\documentclass[superscriptaddress,nofootinbib,aps,prd,10pt]{revtex4-1}
\usepackage{graphicx}
\usepackage{dcolumn}
\usepackage{bm}
\usepackage{amssymb,pslatex}
\usepackage{amsmath,marvosym}
\usepackage{braket}
\renewcommand{\d}{\ensuremath{\mathrm{d}}}

\newcommand{\omu}{\ensuremath{\overline \mu}}

\newcommand{\MSbar}{\overline{\mbox{MS}}}
\newcommand{\lms}{\Lambda_{\overline{\mbox{\tiny{MS}}}}}

\makeatletter
\newsavebox\myboxA
\newsavebox\myboxB
\newlength\mylenA

\newcommand*\xoverline[2][0.75]{%
    \sbox{\myboxA}{$\m@th#2$}%
    \setbox\myboxB\null
    \ht\myboxB=\ht\myboxA%
    \dp\myboxB=\dp\myboxA%
    \wd\myboxB=#1\wd\myboxA
    \sbox\myboxB{$\m@th\overline{\copy\myboxB}$}
    \setlength\mylenA{\the\wd\myboxA}
    \addtolength\mylenA{-\the\wd\myboxB}%
    \ifdim\wd\myboxB<\wd\myboxA%
       \rlap{\hskip 0.5\mylenA\usebox\myboxB}{\usebox\myboxA}%
    \else
        \hskip -0.5\mylenA\rlap{\usebox\myboxA}{\hskip 0.5\mylenA\usebox\myboxB}%
    \fi}
\makeatother

\begin{document}

\title{{\Large {\bf Modeling the Landau-Gauge Ghost Propagator \\[2mm]
                    in 2, 3 and 4 Space-Time Dimensions}}}
\date{\today}

\author{Attilio~Cucchieri}
\email{attilio@ifsc.usp.br}
\affiliation{Instituto de F\'\i sica de S\~ao Carlos, Universidade de S\~ao
             Paulo, Caixa Postal 369, 13560-970, S\~ao Carlos, SP, Brazil}
\author{David~Dudal}
\email{david.dudal@kuleuven.be}
\affiliation{KU Leuven Campus Kortrijk - KULAK, Department of Physics,
             Etienne Sabbelaan 53, 8500 Kortrijk, Belgium}
\affiliation{Ghent University, Department of Physics and Astronomy,
             Krijgslaan 281-S9, 9000 Gent, Belgium}
\author{Tereza~Mendes}
\email{mendes@ifsc.usp.br}
\affiliation{Instituto de F\'\i sica de S\~ao Carlos, Universidade de S\~ao
             Paulo, Caixa Postal 369, 13560-970, S\~ao Carlos, SP, Brazil}
\author{Nele~Vandersickel$\,$}
\email{nele.vandersickel@ugent.be}
\affiliation{Ghent University, Department of Physics and Astronomy,
             Krijgslaan 281-S9, 9000 Gent, Belgium}

\begin{abstract}
We present an analytic description of numerical results for the
ghost propagator $G(p^2)$ in minimal Landau gauge on the lattice.
The data were
produced in the SU(2) case using the largest lattice volumes to date,
for $d = 2, 3$ and 4 space-time dimensions.
Our proposed form for $G(p^2)$ is derived from the one-loop relation
between ghost and gluon propagators, considering a tree-level
ghost-gluon vertex and our previously obtained gluon-propagator
results \cite{Cucchieri:2011ig}.
Although this one-loop expression is not a good description
of the data, it leads to a one-parameter fit of our ghost-propagator
data with a generally good value of $\chi^2/dof$, comparable to other fitting
forms used in the literature.
At the same time, we present a simple parametrization
of the difference between the lattice data and the
one-loop predictions.
\end{abstract}

\maketitle


\section{Introduction}
\label{sec:intro}

An analytic description of propagators and vertices of Yang-Mills
theories ---at the nonperturbative level, in a given gauge--- is a possible
starting point for understanding the relevant features of these theories
and, in particular, the phenomenon of color confinement \cite{Alkofer:2000wg,
Greensite:2011zz,Cloet:2013jya,Brambilla:2014aaa}.
From this point of view, the first natural step is the study of the
infrared (IR) behavior of the gluon propagator $D(p^2)$ and of the
ghost propagator $G(p^2)$ as functions of the momentum $p$.
In the last thirty years, many numerical and analytic studies
have addressed this issue in Landau gauge, in two, three and four
space-time dimensions (see, for example, the reviews
\cite{Brambilla:2014aaa,Binosi:2009qm,Cucchieri:2010xr,Boucaud:2011ug,
Maas:2011se,Vandersickel:2012tz} and references therein).
All the numerical studies, usually done for pure SU(2) and SU(3) lattice gauge
theories, now agree that, in three and in four space-time dimensions
\cite{Bogolubsky:2007ud,Cucchieri:2007md,Sternbeck:2007ug,Cucchieri:2007rg,
Cucchieri:2008fc,Bogolubsky:2009dc,Bornyakov:2009ug,Cucchieri:2011ig,
Cucchieri:2012gb,Oliveira:2012eh,Bornyakov:2013pha,Bornyakov:2013ysa}, the
gluon propagator is IR-finite and the ghost propagator is free-like in the
same limit.
On the contrary, in the $2d$ case \cite{Maas:2007uv,
Cucchieri:2007rg,Cucchieri:2008fc,Cucchieri:2011um,Cucchieri:2011ig,
Cucchieri:2012gb}, the gluon propagator goes to zero at small momenta and
the ghost propagator is IR-enhanced.
In the former case, the numerical data can be related to the so-called
massive solution of the Dyson-Schwinger equations \cite{Aguilar:2004sw,
Aguilar:2006gr,Aguilar:2007fe,Aguilar:2008xm,Aguilar:2009ke,Aguilar:2011ux,
Binosi:2012sj,Aguilar:2012rz}, while in the latter case one should refer
instead to the so-called scaling solution of
these equations \cite{Alkofer:2000wg,
von Smekal:1997vx,Zwanziger:2001kw,Lerche:2002ep,Fischer:2006ub,Huber:2007kc,
Alkofer:2008jy,Huber:2012zj}.
The two different types of solutions can also be related
to the Gribov-Zwanziger (GZ) \cite{Gribov:1977wm,Zwanziger:1989mf,
Zwanziger:1990by,Zwanziger:1991gz,Zwanziger:1992qr,Zwanziger:1993dh,
Capri:2012wx}
and Refined GZ (RGZ) \cite{Dudal:2008sp,Dudal:2008rm,Dudal:2008xd,
Sorella:2011tu,Vandersickel:2011ye,Vandersickel:2011zc,Dudal:2011gd}
approaches, which correspond respectively to the scaling and to the
massive behaviors for the gluon and ghost propagators in the deep IR
limit.\footnote{The interested reader should see Refs.\ \cite{Frasca:2007uz,
Kondo:2009ug,Tissier:2010ts,Tissier:2011ey,Weber:2011nw,Pennington:2011xs,
LlanesEstrada:2012my,Weber:2014lxa,Pelaez:2014mxa,Siringo:2015aka,Machado:2016cij} for
other approaches and points of view on the scaling and/or the massive
solutions.}

In Refs.\ \cite{Cucchieri:2011ig,Cucchieri:2012gb} we have presented an
analytic description of lattice data
\cite{Cucchieri:2007md,Cucchieri:2007rg}
for the SU(2) Landau-gauge gluon
propagator $D(p^2)$ in two, three and four space-time dimensions $d$.
For the cases $d = 3$ and 4, the numerical data can be well fitted using
tree-level predictions of the RGZ approach, i.e.\ considering sums of
propagators of the type $\alpha / (p^2 + \omega^2)$, where $\alpha$
and $\omega$ are in general complex constants.\footnote{Let us mention
that this proposed behavior for the gluon propagator, i.e.\ a pole
structure with complex-conjugate masses (with comparable real and
imaginary parts), can be interpreted as describing an unstable particle.
This is discussed in Ref.\ \cite{Cucchieri:2011ig}, where we also compute
the resulting mass and decay width for the gluon in the $4d$ case.}
On the contrary, in the $2d$ case, no such predictions are
available \cite{Dudal:2008xd}, and the data may be fitted using
a noninteger power of $p$ in the numerator of $D(p^2)$.
These fitting forms have subsequently been used in
Ref.\ \cite{Cucchieri:2012cb}
to evaluate the one-loop-corrected ghost propagator $G(p^2)$ and to
analyze the behavior of the so-called Gribov ghost form factor
$\sigma(p^2)$, defined by
\begin{equation}
G(p^2) \, = \, \frac{1}{p^2} \, \frac{1}{1 - \sigma(p^2)} \,\; ,
\label{eq:Gsigma}
\end{equation}
i.e.,
\begin{equation}
\sigma(p^2) \, \equiv \, 1 - \left[p^2\,G(p^2)\right]^{-1} \; .
\label{eq:sigmabis}
\end{equation}
Using these analytic results one can show that,
considering the bare coupling constant
$g^2$ as a free parameter, the massive solution $G(p^2) \sim 1/p^2$,
corresponding to $\sigma(0) < 1$, is obtained for all values of $g^2$
smaller than a ``critical'' value $g^2_c$.
At $g^2_c$, one has $\sigma(0) = 1$ and the
ghost propagator is IR-enhanced.
These findings confirm that, in the Dyson-Schwinger-equation approach,
the ghost propagator admits a one-parameter
family of behaviors \cite{Boucaud:2008ji,Boucaud:2008ky,Fischer:2008uz,
RodriguezQuintero:2010wy}, labeled by the coupling constant $g^2$.

In this work we present the final step of our analysis, using the
one-loop results for $G(p^2)$ of Ref.\ \cite{Cucchieri:2012cb} as
theoretical predictions for the analytic modeling of numerical data
\cite{Cucchieri:2007md,Cucchieri:2008fc}
for the ghost propagator in Landau gauge in two, three and four space-time
dimensions.
(Similar studies have been presented in Refs.\ \cite{Dudal:2012zx,
Aguilar:2013xqa,Aguilar:2014rva} for the four-dimensional case.)
We find that the proposed analytic forms {\em do not} yield a good
description of the ghost-propagator data.
This is in agreement with Refs.\ \cite{Dudal:2012zx,Aguilar:2013xqa,
Aguilar:2014rva}.
Nevertheless, by treating $g^2$ as a free parameter in these forms,
one obtains fits of $G(p^2)$ with generally good values of $\chi^2/dof$,
comparable to other fitting forms used in the literature (see e.g.\
\cite{Cucchieri:2008fc,Aguilar:2008xm}).
Finally, we attempt a simple parametrization of the
difference between the lattice data and the one-loop predictions,
which turns out to be very similar for the $d= 2, 3$ and $4$ cases.
This supports a possible
interpretation of the physical effects that are missing in the one-loop
results \cite{Dudal:2012zx,Aguilar:2013xqa,Aguilar:2014rva,Huber:2012kd}.

The paper is organized as follows.
In the next section, we recall the main results of Refs.\
\cite{Cucchieri:2011ig,Cucchieri:2012gb,Cucchieri:2012cb} and, in
particular, the formulae used in our analysis of the ghost propagator.
Then, in Section \ref{sec:fits}, we present and discuss the fits to
the lattice data.
Lastly, in Section \ref{sec:conclusions}, we outline our conclusions.


\section{One-Loop Predictions}
\label{sec:RGZ}

As already explained in the Introduction, in Refs.\ \cite{Cucchieri:2011ig,
Cucchieri:2012gb} the SU(2) gluon propagator has been fitted in $d = 3, 4$
and 2 using, respectively, the functions\footnote{Note that, here and in
the following, we choose to present results for the $3d$ and $4d$ cases
before the $2d$ case.}
\begin{equation}
D(p^2) \, = \, C \,
    \frac{(p^2 + s) \, (p^2 + 1)}{(p^4 + u^2 \, p^2 + t^2) \, (p^2 + k)}
        \; ,
\label{eq:fDvrgzsimple}
\end{equation}
\begin{equation}
D(p^2) \, = \, C \, \frac{p^2 + s}{p^4 + u^2 \, p^2 + t^2} \;
\label{eq:f4dgluon}
\end{equation}
and
\begin{equation}
D(p^2) \, = \, C \,
          \frac{p^2 + l\,p^{\eta} + s}{p^4 + u^2 \, p^2 + t^2} \; .
\label{eq:f2dgluon}
\end{equation}
The first two propagators are tree-level expressions obtained in the
RGZ approach \cite{Dudal:2008sp,Dudal:2008rm,Dudal:2008xd,Sorella:2011tu,
Vandersickel:2011ye,Vandersickel:2011zc,Dudal:2011gd}, while
the last formula is a simple generalization of the form
in Eq.\ (\ref{eq:f4dgluon}).
Note that these three functions can be expressed as linear combinations of
propagators of the type $\, 1 / (p^2+\omega^2)$, where $\omega^2$ is in general
a complex number.
In particular, Eqs.\ (\ref{eq:fDvrgzsimple}) and (\ref{eq:f4dgluon}) can be
re-written respectively as
\begin{equation}
D(p^2) \, = \, \frac{\alpha}{p^2+\omega_1^2} + \frac{\beta}{p^2+\omega_2^2}
                             + \frac{\gamma}{p^2+\omega_3^2}
\label{gluonpropsimp}
\end{equation}
and
\begin{equation}
D(p^2) \, = \, \frac{\alpha_+}{p^2+\omega_{+}^2}
                       + \frac{\alpha_-}{p^2+\omega_{-}^2} \; .
\label{4D3}
\end{equation}
The fits to the data \cite{Cucchieri:2011ig,Cucchieri:2012gb} suggest that,
in the $3d$ case [see Eq.\ (\ref{gluonpropsimp})], one root is real,
for example $\omega_1$, while the
other two roots are complex-conjugate, i.e.\ $(\omega_2^{2})^{*} =
\omega_3^2$, implying also $\beta = \gamma^{*}$.
Similarly, in the $4d$ case [see Eq.\ (\ref{4D3})] one finds,
by fitting the lattice data, that
$\omega_{\pm}^2$ are complex-conjugate roots, i.e.\ $\omega_{-}^2 =
(\omega_{+}^2)^*$ and $\alpha_- = \alpha_+^*$.
On the other hand, in the $2d$ case we need to consider the more general
form $ p^{\eta} / (p^2+\omega^2)$ with $\eta \geq 0$.
Indeed, one can re-write Eq.\ (\ref{eq:f2dgluon}) as
\begin{equation}
D(p^2) \, = \,
\frac{\alpha_{+} \, + \, i c p^{\eta}}{p^2 + \omega_+^2} \, + \,
\frac{\alpha_{-} \, - \, i c p^{\eta}}{p^2 + \omega_-^2} \; ,
\label{eq:D2D}
\end{equation}
where $c$ is real, $\alpha_{-} = \alpha_{+}^*$ and $\omega_{-}^2 =
(\omega_{+}^2)^*$. Estimates for the fitting parameters of the functions
(\ref{eq:fDvrgzsimple})--(\ref{eq:D2D}) can be found, respectively, in Tables
IX, II, XIII, XI, IV and XIV of Ref.\ \cite{Cucchieri:2011ig}.

Using the notation of Ref.\ \cite{Cucchieri:2012cb}, the one-loop-corrected
Landau-gauge ghost propagator can be evaluated [for the SU($N_c$) gauge
group in the $d$-dimensional case] using the relation
\begin{equation}
G(p^2) \; = \; \frac{1}{p^2}
       \; - \; \frac{\delta^{ab}}{N_c^2 - 1} \,
       \frac{1}{p^4} \, g^2 f^{adc} f^{cdb} \, \int \frac{\d^d q}{(2\pi)^d}
       \, (p - q)_{\mu} \; p_{\nu} \; D(q^2) \, P_{\mu \nu}(q)
       \, \frac{1}{(p - q)^2} \; ,
\label{eq:Gini}
\end{equation}
where $ \delta^{ab} \, D(q^2) \, P_{\mu\nu}(q) $ stands for the
gluon-propagator forms described above and
$P_{\mu\nu}(q) = \left(\delta_{\mu\nu}-q_\mu q_\nu/q^2\right)$
is the usual projector onto the transverse sub-space. Here we have
considered the tree-level ghost-gluon vertex $\; i g f^{adc} p_{\nu} $,
where $p$ is the outgoing ghost momentum. The color indices $a, d, c$ refer,
respectively, to the incoming ghost, to the gluon and to the outcoming
ghost. Then, using for the gluon propagator the expressions
(\ref{gluonpropsimp})--(\ref{eq:D2D}) above and writing $G(p^2)$ as in
Eq.\ (\ref{eq:Gsigma}), one can show \cite{Cucchieri:2012cb} that the
Gribov ghost form factor $\sigma(p^2)$ is given in three, four and two
space-time dimensions by the formulae reported in the subsections below.

\subsection{The three-dimensional case}
\label{sec:3dtheory}

Assuming (see above) that $\omega_1$ and $\alpha$ are real and
writing the remaining fitting parameters of Eq.\ (\ref{gluonpropsimp}) as
\begin{equation}
    \beta  \, = \, a + i b \; , \qquad
    \gamma \, = \, a - i b
\end{equation}
and
\begin{equation}
\omega_2^2 \, = \, v + i w \; , \qquad
\omega_3^2 \, = \, v - i w \; ,
\end{equation}
we obtain \cite{Cucchieri:2012cb}
\begin{equation}
\sigma_{1L}(p^2) \; = \; \frac{g^2 N_c}{8}
    \left[ \frac{\alpha\; s(p^2,\omega_1^2)}{4\,\pi\,\omega_1^2\,p^3}
                    \, + \, f_R(p^2) \right]\,,
\label{eq:sigma3d2}
\end{equation}
where
\begin{eqnarray}
s(p^2,\omega^2) & = & -\pi \, p^4
              \,+ \, 2 \, p^3 \, \sqrt{\omega^2} \,- \,
      2 \, p \, (\omega^2)^{3/2} \, + \, 2 \,
   (p^2 \, + \, \omega^2)^2 \,
      \arctan\left(\frac{p}{\sqrt{\omega^2}}\right) \; , \\[3mm]
f_R(p^2) & = &
  f_1(p^2) \, + \, f_2(p^2) \, + \, f_3(p^2) \,
                        + \, f_4(p^2) \, + \, f_5(p^2)
     \label{eq:f3d}
\end{eqnarray}
with
\begin{eqnarray}
R      & = & \sqrt{v^2+w^2} \; , \label{eq:Rdef} \\[2mm]
f_1(p^2) & = & - p \; \frac{a v \, + \, b w}{2 \, R^2} \; , \\[2mm]
f_2(p^2) & = & \frac{\left( a v \, + \, b w \right) \sqrt{R+v} \,-\,
   \left( b v \, - \, a w \right) \sqrt{R-v}}{\sqrt{2} \, \pi \, R^2}
                                    \; , \\[2mm]
f_3(p^2) & = & - \frac{1}{p^2} \,
      \frac{a \sqrt{R+v} \,-\,
             b \sqrt{R-v}}{\sqrt{2} \, \pi} \; , \\[2mm]
f_4(p^2) & = & A(p^2) \;
        \frac{p^4 \left( a v \, + \, b w \right) \,+\, 2\,a\,p^2\,R^2
   \,+\,R^2 \left( a v \, - \, b w \right)}{2 \, \pi \, R^2\, p^3}
                              \; , \\[2mm]
f_5(p^2) & = &-L(p^2) \;
      \frac{p^4 \left( b v \, - \, a w \right) \,+\, 2\,b\,p^2\,R^2
        \,+\,R^2 \left( b v \, + \, a w \right)}{2 \, \pi \, R^2\, p^3}
\end{eqnarray}
and
\begin{eqnarray}
A(p^2) & = & \left\{ \begin{array}{ll}
  \arctan\left( \frac{\sqrt{2} \, p \,
           \sqrt{R+v}}{R\,-\,p^2} \right) \qquad
            & \mbox{if} \;\;\; R\,-\,p^2 > 0 \\[3mm]
  \pi \,+\, \arctan\left( \frac{\sqrt{2} \, p \,
           \sqrt{R+v}}{R\,-\,p^2} \right) \qquad
             & \mbox{if} \;\;\; R\,-\,p^2 < 0
          \end{array} \right. \; ,\\[3mm]
L(p^2) & = & \ln\left[ \frac{\sqrt{p^4 \,+\, 2 \, p^2\,v \,+\, R^2}}{
     R \,+\, p \, \left( p \,+\, \sqrt{2}\,\sqrt{R-v} \right)} \right]
                \label{eq:r3d}  \; .
\end{eqnarray}

\subsection{The four-dimensional case}
\label{sec:4dtheory}

By working in the $\MSbar$ scheme, using dimensional regularization
and writing the fitting parameters of Eq.\ (\ref{4D3}) as $\alpha_{\pm} =
a \pm i b $ and $\omega_{\pm}^2 = v \pm i w $, one finds \cite{Cucchieri:2012cb}
\begin{equation}
\sigma^{\overline{\mbox{\tiny{MS}}}}_{1L}(p^2) \, = \,
                   \frac{g^2 N_c}{32 \pi^2 R^2}
        \left[ -p^2 t_1(p^2) \, + \, R^2 t_2(p^2) \, + \,
       p^{-2} t_3(p^2) \, - \,
             p^{-4} t_4(p^2) \right] \label{eq:sigma4d2}
\end{equation}
with $R$ defined in Eq.\ (\ref{eq:Rdef}),
\begin{eqnarray}
t_1(p^2) & = & (a v+b w) [\ell_2(p^2)+\ell_3(p^2)]
                  \,-\, (b v-a w) [a_1(p^2)-a_2(p^2)] \; , \\[2mm]
t_2(p^2) & = & a [5+\ell_1(p^2)+\ell_2(p^2)+\ell_3(p^2)-4 \ell_4(p^2)]
               \, - \, b[a_1(p^2)-a_2(p^2) - 4 a_3(p^2)] \; , \\[2mm]
t_3(p^2) & = & [1-3 \ell_3(p^2)] (a v^3-b w v^2+v a w^2-b w^3)
            \,- \, 3 a_2(p^2) (b v^3+a w v^2+v b w^2+a w^3) \; , \\[2mm]
t_4(p^2) & = & \ell_3(p^2) (a v^4-2 w b v^3-2 v b w^3-a w^4)
               \, + \, a_2(p^2) (b v^4+2 a w v^3+2 v a w^3-b w^4)
\end{eqnarray}
and
\begin{eqnarray}
\ell_1(p^2) & = & \ln\left(\frac{p^2}{\omu^2}\right) \; , \\[2mm]
\ell_2(p^2) & = & \ln\left(\frac{R}{p^2}\right) \; , \\[2mm]
\ell_3(p^2) & = & \ln\left( \frac{\sqrt{R^2 p^4+R^4+2 v R^2 p^2}}{R^2}
                          \right) \; , \label{eq:l2} \\[2mm]
\ell_4(p^2) & = & \ln\left(
         \frac{\sqrt{p^4+2 v p^2+R^2}}{\omu^2}\right) \; , \\[2mm]
a_1(p^2)    & = & \arctan\left( \frac{w}{v} \right) \; , \\[2mm]
a_2(p^2) & = & \arctan\left( \frac{w p^2}{R^2+v p^2}\right)
                       \; , \label{eq:a2} \\[2mm]
a_3(p^2) & = & \arctan\left( \frac{w}{v+p^2}\right)
                       \; . \label{eq:a3}
\end{eqnarray}

The above result for $\sigma^{\overline{\mbox{\tiny{MS}}}}_{1L}(p^2)$ cannot,
however, be directly compared to the lattice data, since the $\MSbar$ scheme
is defined only at the perturbative level.
Thus, in order to make this comparison in the next section, we use a
momentum-subtraction (MOM) renormalization scheme defined by
\begin{equation}
\label{mom}
\left. D^{\mbox{\tiny{MOM}}}(p^2) \right|_{p^2=\omu^2} \, = \,
             \frac{1}{\omu^2} \; , \qquad
\left. G^{\mbox{\tiny{MOM}}}(p^2) \right|_{p^2=\omu^2} \, = \,
             \frac{1}{\omu^2} \; .
\end{equation}
The MOM-scheme condition for the gluon propagator affects only the global
multiplicative factor $C$ in Eq.\ (\ref{eq:f4dgluon}), or the parameters
$\alpha_{\pm}$ in Eq.\ (\ref{4D3}). As a consequence [see Eqs.\
(\ref{eq:sigma4d2})--(\ref{eq:a3})] the quantity $\sigma^{\overline{
\mbox{\tiny{MS}}}}_{1L}(p^2)$ also gets modified by a global factor. At the
same time, we can transform the above $\MSbar$ result for $G(p^2)$ into
the MOM scheme by writing\footnote{This corresponds to a one-loop
(finite) shift in the renormalization factor of the ghost propagator.}
\begin{equation}
G^{\mbox{\tiny{MOM}}}(p^2) \, = \, \frac{1}{p^2} \, \left[ \,
 1 \, - \, \sigma^{\overline{\mbox{\tiny{MS}}}}_{1L}(p^2) \, + \,
                h(\omu^2)\, \right]^{-1} \; ,
\label{eq:G4d}
\end{equation}
where the parameter $h(\omu^2)$ is fixed by imposing the MOM-scheme condition (\ref{mom}),
i.e.\
\begin{equation}
\sigma^{\overline{\mbox{\tiny{MS}}}}_{1L}(\omu^2) \, = \, h(\omu^2) \; .
\end{equation}

\subsection{The two-dimensional case}
\label{sec:2dtheory}

In the $2d$ case one finds \cite{Cucchieri:2012cb}
\begin{equation}
\sigma_{2d}(p^2) \, = \,
     g^2 N_c \, \left[ \, \alpha_+ \, f(p^2,\omega_+^2) \, + \,
                          \alpha_- \, f(p^2,\omega_-^2)
     \, + \, i c \widetilde{f}(p^2,\omega_+^2,\eta) \, - \,
     i c \widetilde{f}(p^2,\omega_-^2,\eta) \,\right]
\label{eq:sigma2d}
\end{equation}
with
\begin{equation}
f(p, \omega^2) \, = \,
   \frac{1}{8 \pi} \left[ \, \frac{1}{p^2} \,
          \ln\left(1 + \frac{p^2}{\omega^2}\right) \, + \,
     \frac{1}{\omega^2} \, \ln\left(1 + \frac{\omega^2}{p^2}\right)
                  \, \right]
\label{eq:fwithe0}
\end{equation}
and
\begin{equation}
\widetilde{f}(p, \omega^2, \eta) \, = \,
       \frac{(\omega^2)^{\eta/2}}{4 \pi \, \eta \, p^2} \,
      \left[ \frac{p^2+\omega^2}{\omega^2} \,
     B\left(\frac{\omega^2}{p^2+\omega^2}; 1-\frac{\eta}{2} ,
         1+\frac{\eta}{2} \right)
    \, - \, B\left(1-\frac{\eta}{2} , 1+\frac{\eta}{2} \right)
                \right] \; .
\label{eq:fwithe}
\end{equation}
Here,
\begin{equation}
B\left(x;a,b\right) \, = \,
     \int_0^x \, \d t \, t^{a-1} \, \left( 1 - t \right)^{b-1}
\label{eq:Betainc}
\end{equation}
is the incomplete Beta function, which is defined for $a, b > 0$
\cite{GR}, implying $2 > \eta$ in our case, and $B(a,b) \equiv
B(1;a,b)$ is the Beta function.
Then, by writing $\alpha_{\pm} = a \pm i b $ and $\omega_{\pm}^2
= v \pm i w $ one gets for the first two terms of Eq.\
(\ref{eq:sigma2d}) above the expression
\begin{eqnarray}
\alpha_+ \, f(p^2,\omega_+^2) \, + \,
\alpha_- \, f(p^2,\omega_-^2) \;& = &\, \frac{1}{8 \pi} \, \left\{ \,
    \frac{1}{p^2} \, \left[ a \, \ell_3(p^2) \, + \,
          b \, a_2(p^2) \right] \right. \nonumber \\[2mm]
& & \left. \qquad \, + \, \frac{1}{R^2} \,
         \left[ \left(a v + b w\right) \ell_5(p^2)
   \, - \, \left(b v - a w\right) a_3(p^2) \right] \, \right\} \; ,
\label{eq:sigmareal2d}
\end{eqnarray}
where
\begin{equation}
\ell_5(p^2) \, = \,
  \ln\left( \frac{\sqrt{p^4+2 v p^2+R^2}}{p^2}\right) \;
\end{equation}
and $R$, $\ell_3(p^2)$, $a_2(p^2)$ and $a_3(p^2)$ have already
been defined in Eqs.\ (\ref{eq:Rdef}), (\ref{eq:l2}), (\ref{eq:a2})
and (\ref{eq:a3}). We also have
\begin{equation}
i c \widetilde{f}(p^2,\omega_+^2,\eta) \, - \,
i c \widetilde{f}(p^2,\omega_-^2,\eta) \, = \, - 2 \, c \, \Im
   \left[ \, \widetilde{f}(p^2,\omega_+^2,\eta) \, \right] \; ,
\label{eq:imaginaryw}
\end{equation}
where we have indicated with $\Im$ the imaginary part of the expression
in square brackets.


\begin{table}[b]
\protect\vskip 1mm
\begin{tabular}{| c | c | c | c | c | c | c |}
\hline
$V=N^d$   &   $\beta$   & $\#$ confs & $a \, (fermi)$
          & $L=N a \, (fermi)$ & $p_{min} \, (\text{MeV})$ & $r$
\\ \hline
\hline
$140^3$  &  $3.0$  & 626  & 0.268  &  37.5  &  33.0 & 0.018  \\ \hline
$200^3$  &  $3.0$  & 484  & 0.268  &  53.6  &  23.1 & 0.006  \\ \hline
$240^3$  &  $3.0$  & 343  & 0.268  &  64.3  &  19.2 & 0.000  \\ \hline
$320^3$  &  $3.0$  & 122  & 0.268  &  85.8  &  14.4 & 0.012  \\ \hline
\hline
$48^4$   &  $2.2$  &  99  & 0.210  &  10.1  & 122.7 & 0.017  \\ \hline
$56^4$   &  $2.2$  & 100  & 0.210  &  11.8  & 105.2 & 0.007  \\ \hline
$64^4$   &  $2.2$  & 100  & 0.210  &  13.4  &  92.1 & 0.047  \\ \hline
$80^4$   &  $2.2$  &  97  & 0.210  &  16.8  &  73.7 & 0.021  \\ \hline
$128^4$  &  $2.2$  &  21  & 0.210  &  26.9  &  46.0 & 0.012  \\ \hline
\hline
$80^2$   & $10.0$  & 600  & 0.219  &  17.5  &  70.6 & 0.006  \\ \hline
$120^2$  & $10.0$  & 600  & 0.219  &  26.3  &  47.1 & 0.005  \\ \hline
$160^2$  & $10.0$  & 600  & 0.219  &  35.0  &  35.3 & 0.008  \\ \hline
$200^2$  & $10.0$  & 600  & 0.219  &  43.8  &  28.3 & 0.001  \\ \hline
$240^2$  & $10.0$  & 600  & 0.219  &  52.6  &  23.5 & 0.015  \\ \hline
$280^2$  & $10.0$  & 600  & 0.219  &  61.3  &  20.2 & 0.000  \\ \hline
$320^2$  & $10.0$  & 600  & 0.219  &  70.1  &  17.7 & 0.008  \\ \hline
\end{tabular}
\caption{For each lattice volume $V$ and lattice coupling $\beta$
we indicate the number of configurations considered, the value
of the lattice spacing $a$ in $fermi$, the lattice size
$L$ (also in $fermi$), the value of the smallest
nonzero momentum $p_{min} =  2 \sin(\pi / N) / a$ (in
$\text{MeV}$) and the coefficient $r$ that allows the largest reduction
of the rotational-symmetry-breaking effects [see Eq.\ (\ref{eq:pimpr})
and explanation in the text].
\label{tab:lattice}}
\end{table}

\section{Fits to Ghost-Propagator Data}
\label{sec:fits}

The data for the ghost propagators $G(p^2)$ in $d = 3, 4$ and $2$ have been
evaluated for essentially the same set of lattice parameters considered
for the gluon propagator $D(p^2)$ in Refs.\ \cite{Cucchieri:2011ig,
Cucchieri:2012gb}.
A summary of the various lattice setups is presented in Table
\ref{tab:lattice}.
More details about the numerical simulations can be found in Ref.\
\cite{Cucchieri:2011ig}.
These simulations \cite{Cucchieri:2007md,Cucchieri:2007rg,Cucchieri:2008fc}
have been done in 2007 using, in $3d$ and in $4d$, the
4.5 Tflops IBM supercomputer at LCCA--USP 
and, in the $2d$ case, a PC cluster at the IFSC--USP.
In all cases we set the lattice spacing $a$ by relating the lattice string
tension $\sqrt{\sigma_{latt}}$ to the physical value
$\sqrt{\sigma} \approx 0.44 \, \text{GeV}$, which
is a typical value for this quantity in the $4d$ SU(3) case.
For $\sqrt{\sigma_{latt}}$ we used the results described in
\cite{Cucchieri:2003di}, \cite{Bloch:2003sk} and \cite{Cucchieri:1995pn},
respectively for $d = $ 3, 4 and 2.
Note that all runs are in the scaling region and all data refer to the
SU(2) case.
Possible systematic effects due to Gribov copies \cite{Cucchieri:1997dx,
Silva:2004bv,Bogolubsky:2005wf,Bogolubsky:2009qb,Maas:2009ph,
Bornyakov:2009ug,Sternbeck:2012mf} or unquenching effects
\cite{Boucaud:2001un,Furui:2005bu,Ilgenfritz:2006he,Bowman:2007du,
Silva:2010vx} were {\em not} considered.

Let us also recall that the Landau-gauge ghost propagator $G(p^2)$
is obtained by inverting the lattice Faddeev-Popov matrix
${\cal M}(b,x;c,y)$ and is given by
\begin{equation}
G^{bc}(p^2)\, = \,\sum_{x\mbox{,}\, y}
\frac{e^{- 2 \pi i \, \hat{p} \cdot (x - y) / N}}{V}\,
\langle\, {\cal M}^{- 1}(b,x;c,y) \,\rangle \;=\;
\,\delta^{bc}\, G(p^2) \; ,
\label{eq:Gdef}
\end{equation}
where $b$ and $c$ are color indices and $\langle \;\;\rangle$ stands for
the path-integral average.
The inversion of the Faddeev-Popov matrix is obtained by using a
conjugate gradient method with even/odd preconditioning and
point sources \cite{Boucaud:2005gg,Cucchieri:2006tf}.
For the lattice Faddeev-Popov matrix we consider Eq.\ (22) in
Ref.\ \cite{Cucchieri:2005yr}.
At the same time, the momentum components $p_{\mu}$ are given by
\begin{equation}
p_{\mu} \;=\; 2 \, \sin\left(\frac{\pi \, \hat{p}_{\mu}}{N}\right)
\label{eq:pmu}
\end{equation}
and $\hat{p}_{\mu}$ takes the values $0, 1, \ldots, N-1$. However, since the
Faddeev-Popov matrix has a trivial null eigenvalue corresponding
to a constant eigenvector, one cannot evaluate the ghost propagator
at zero momentum, i.e.\ with $\hat{p}_{\mu} = 0$ for all directions
$\mu$.
For the nonzero momenta, we considered in $2d$ all momenta with
components $(\xoverline{p},0)$ and $(\xoverline{p},\xoverline{p})$, plus all
possible permutations of the components. Similarly, in $3d$ and in
$4d$ we present results for momenta of the type $(\xoverline{p},0,0)$,
$(\xoverline{p},\xoverline{p},0)$, $(\xoverline{p},\xoverline{p},\xoverline{p})$ and of
the type $(\xoverline{p},0,0,0)$, $(\xoverline{p},\xoverline{p},0,0)$,
$(\xoverline{p},\xoverline{p},\xoverline{p},0)$ and $(\xoverline{p},\xoverline{p},
\xoverline{p},\xoverline{p})$, respectively.\footnote{Again, for each
kinematic setup, we consider all possible permutations of the
momentum components. When permutations of the momentum components
were available, the average over different permutations was taken
independently for each configuration.}

\begin{figure}
\begin{center}
\includegraphics[width=.48\textwidth]{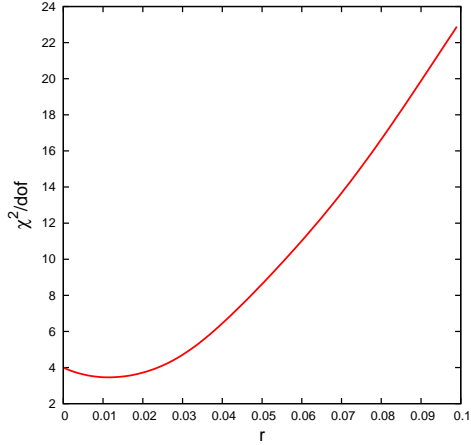}
\caption{Plot of the average reduced chi-squared statistic $\chi^2/dof$
as a function of the parameter $r$ [see Eq.\ (\ref{eq:pimpr}) and
explanation in the text]. The average is taken over all lattice volumes
$V$ and $\beta$ values considered.}
\label{fig:chireduced}
\end{center}
\end{figure}

\begin{figure}
\begin{center}
\includegraphics[width=.48\textwidth]{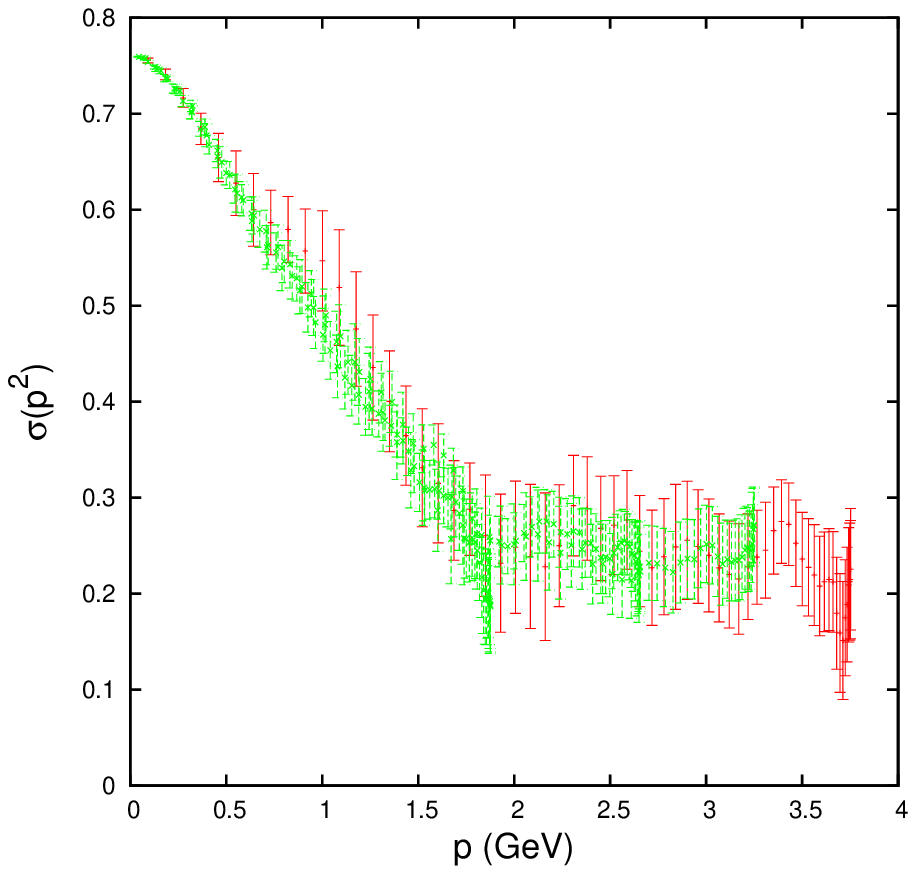}
\hskip 5mm
\includegraphics[width=.48\textwidth]{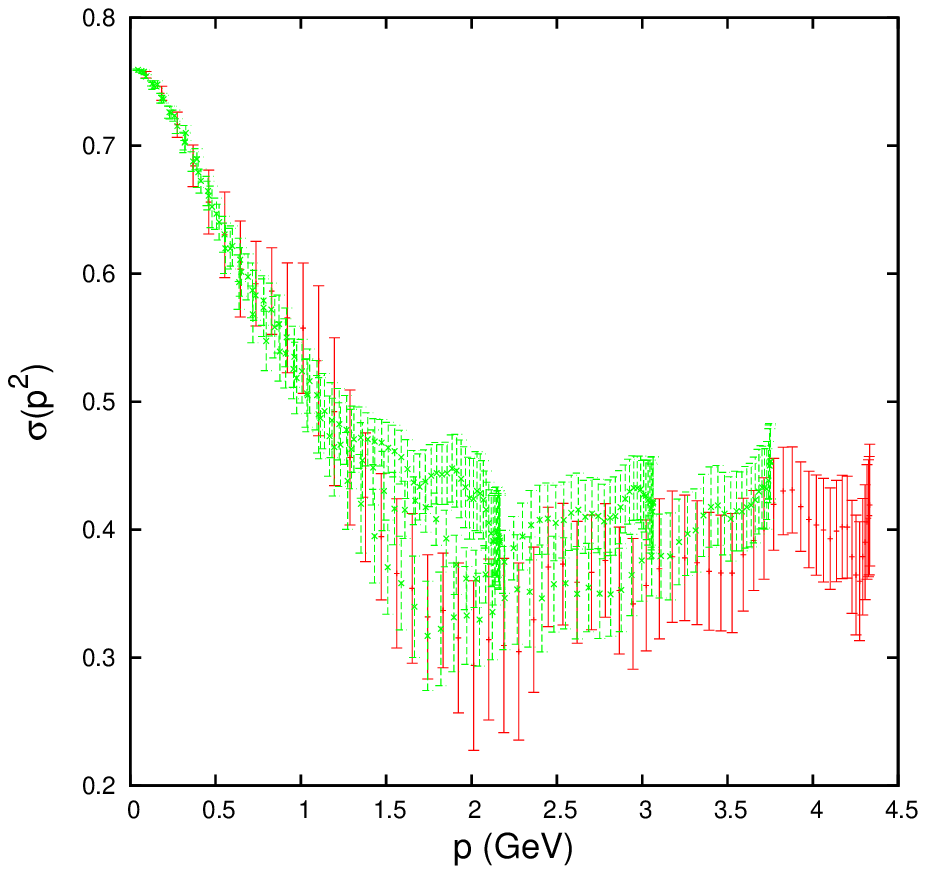}
\caption{Plot of the Gribov ghost form factor $\sigma(p^2)$ [see Eq.\
(\ref{eq:sigmabis})] for the lattice volume $V = 128^4$ at $\beta = 2.2$,
as a function of unimproved momenta [see Eq.\ (\ref{eq:punimpr})]
(left plot) and of improved momenta [see Eq.\ (\ref{eq:pimpr})]
with $r=1/12$ (right plot). In both plots, red data points correspond to
momenta along the diagonal direction ($p_{\mu} = \xoverline{p}$ for
$\mu = 1,\ldots,4$), while green data points correspond to
off-diagonal momenta. All momenta are in physical units. Error bars have been
estimated using propagation of errors.
}
\label{fig:rotat}
\end{center}
\end{figure}

Finally, we recall that the best fits for the gluon propagator data,
reported in Refs.\ \cite{Cucchieri:2011ig,Cucchieri:2012gb} and
used here as theoretical inputs, were obtained by considering this
propagator as a function of the improved magnitude squared of the momentum
\cite{Ma:1999kn}
\begin{equation}
p^2 \, = \, \sum_{\mu} \, p_{\mu}^2 \, + \, r \,
                  \sum_{\mu} \, p_{\mu}^4 \; ,
\label{eq:pimpr}
\end{equation}
with $r = 1/12 \approx 0.083$. This allows a better control of systematic
effects ---related to the breaking of rotational symmetry
\cite{Ma:1999kn,Leinweber:1998uu,deSoto:2007ht}--- than the usual
unimproved definition
\begin{equation}
p^2 \, = \, \sum_{\mu} \, p_{\mu}^2 \; .
\label{eq:punimpr}
\end{equation}
On the other hand, for the ghost propagator, the data are generally smoother
when using the above unimproved definition,\footnote{This is probably related
to the fact that the ghost propagator $G(p^2)$ [see Eq.\ (\ref{eq:Gdef})]
does not depend explicitly on the Lorentz index $\mu$.} or a very small value
of $r$. In order to verify this, we have considered the momentum behavior of
the Gribov ghost form factor $\sigma(p^2)$
[defined in Eq.\ (\ref{eq:sigmabis})]
as a function of the improved magnitude squared of the momenta (\ref{eq:pimpr})
for 100 different values of the parameter $r$,
i.e.\ $r = 0, 0.001, 0.002, 0.003,
\ldots, 0.099$. For each of these values, we used a cubic spline interpolation
to obtain a description of the ghost-propagator data along the diagonal momentum
direction,\footnote{This direction is usually less affected by
rotational-symmetry-breaking effects \cite{Leinweber:1998uu,deSoto:2007ht}.}
i.e.\ for $p_{\mu} = \xoverline{p}$ and $\mu = 1, \ldots, d$. Then, we have evaluated
the goodness of the fit, i.e.\ the reduced chi-squared statistic $\chi^2/dof$,
by comparing this interpolated curve
with ghost-propagator data off the diagonal-momentum
direction, i.e.\ with at least one momentum component equal to zero. In
the last column of Table \ref{tab:lattice} we report, for each lattice
volume $V$ and lattice coupling $\beta$, the value of the parameter
$r$ that yields the smallest value for the reduced chi-squared statistic.
As one can clearly see, these values of $r$ are very small for most
of the cases considered. We also show in Fig.\ \ref{fig:chireduced} the
average value (over all lattice volumes $V$ and $\beta$ values
considered) of $\chi^2/dof$ as a function of the parameter $r$.
Again, we see that for small values of $r$ the $\chi^2/dof$ curve
is almost flat, with a minimum value around $r = 0.01$, and that for
$\, r \geq 0.03$ the average value of the reduced chi-squared increases
almost linearly. The effect of using a large value of the parameter $r$
can also be clearly visualized in the plots reported in Fig.\
\ref{fig:rotat}, where we show the data for $\beta = 2.2$ and our
largest $4d$ lattice $V = 128^4$
as a function of unimproved momenta (left plot) and as a function of
``improved'' momenta\footnote{This value of $r$ is usually employed in
fits of the gluon propagator (see e.g.\ \cite{Cucchieri:2011ig}).}
with $r=1/12$ (right plot). Indeed, the spread of the data
points is clearly larger in the second case. Thus, for simplicity's sake,
we will consider below all the ghost-propagator data as a function of
the unimproved momenta [see Eq.\ (\ref{eq:punimpr})].
One should, of course, try to reduce discretization effects
in order to obtain results closer to the continuum formulation of the
theory, but we must note that
different lattice quantities are subject in general to different such
effects.
Thus, it is not surprising that gluon and ghost propagator data
require different definitions of the lattice momenta when one tries to connect
lattice data to the continuum analysis carried out in Ref.\
\cite{Cucchieri:2012cb}.

\begin{figure}
\begin{center}
\includegraphics[width=.48\textwidth]{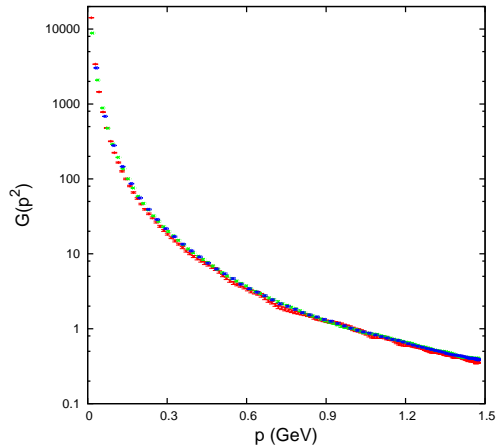}
\caption{Plot of the ghost propagator $G(p^2)$ as a function of the
magnitude of the (unimproved) momenta $p$ (both in physical units)
for the lattice volumes $V = 140^3$ (symbol $*$ in blue), $V = 240^3$
(symbol $\times$ in green) and $V = 320^3$ (symbol $+$ in red) at
$\beta = 3.0$. Here we show the data corresponding to momenta with
only one component different from zero. The data are (multiplicatively)
normalized to 1 for $p=1.0 \;\text{GeV}$.
Notice the logarithmic scale on the $y$ axis.
}
\label{fig:volume3d}
\end{center}
\end{figure}

\begin{figure}
\begin{center}
\includegraphics[width=.48\textwidth]{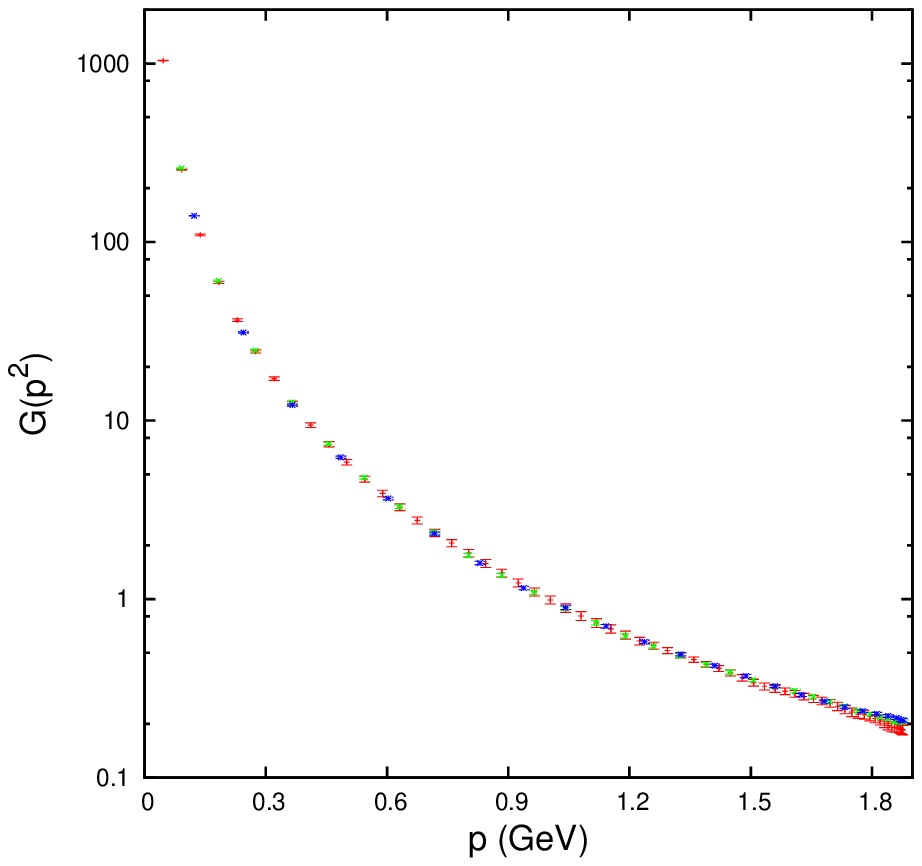}
\caption{Plot of the ghost propagator $G(p^2)$ as a function of the
magnitude of the (unimproved) momenta $p$ (both in physical units)
for the lattice volumes $V = 48^4$ (symbol $*$ in blue), $V = 64^4$
(symbol $\times$ in green) and $V = 128^4$ (symbol $+$ in red) at
$\beta = 2.2$. Here we show the data corresponding to momenta with
only one component different from zero. The data are (multiplicatively)
normalized to 1 for $p=1.0 \;\text{GeV}$.
Notice the logarithmic scale on the $y$ axis.
}
\label{fig:volume4d}
\end{center}
\end{figure}

\begin{figure}
\begin{center}
\includegraphics[width=.48\textwidth]{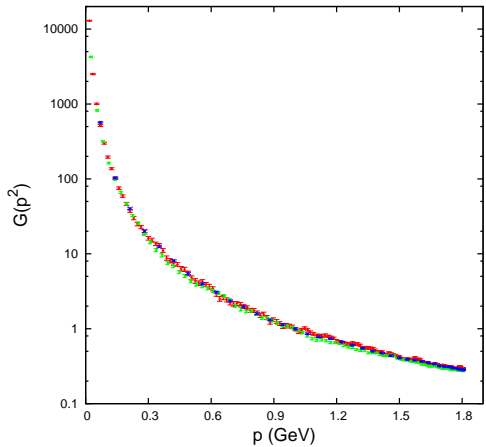}
\caption{Plot of the ghost propagator $G(p^2)$ as a function of the
magnitude of the (unimproved) momenta $p$ (both in physical units)
for the lattice volumes $V = 80^2$ (symbol $*$ in blue), $V = 200^2$
(symbol $\times$ in green) and $V = 320^2$ (symbol $+$ in red) at
$\beta = 10.0$. Here we show the data corresponding to momenta with
only one component different from zero. The data are (multiplicatively)
normalized to 1 for $p=1.0 \;\text{GeV}$.
Notice the logarithmic scale on the $y$ axis.
}
\label{fig:volume2d}
\end{center}
\end{figure}

In the next three subsections we present the modeling of the numerical data
for $G(p^2)$. In analogy with the presentation of the one-loop
calculations in Section \ref{sec:RGZ}, we first give our results for the
$3d$ case, then for the $4d$ case and, finally, for the $2d$ case.
Let us note that finite-size effects for $G(p^2)$ are generally negligible.
This can be seen in Figs.\ \ref{fig:volume3d}, \ref{fig:volume4d}
and \ref{fig:volume2d}, where $G(p^2)$ is plotted for three different
lattice sizes respectively for $d = 3, 4$ and $2$. We also remark that
the use of the point-source method in the evaluation of $G(p^2)$ leads
to the slight ``wiggling'' of the lattice data seen in the three plots
above (see Ref.\ \cite{Cucchieri:2006tf}).
Thus, in the following, we will always use the largest lattice volume
available for each dimension $d$. Also, in all cases we will show the
data (multiplicatively) normalized to
$1/\omu^2$ for $p=\omu=2.5 \,\text{GeV}$.

The analytic expression proposed for the ghost propagator will be cast in the
form of Eq.\ (\ref{eq:Gsigma}), using in each dimension $d$
the corresponding one-loop results \cite{Cucchieri:2012cb}
for the Gribov ghost form factor $\sigma(p^2)$
listed in Section \ref{sec:RGZ} above.
The parameters in $\sigma(p^2)$ will be taken from the
gluon-propagator results obtained in
Refs.\ \cite{Cucchieri:2011ig,Cucchieri:2012gb}.
Then, the only parameter left is the bare coupling constant $g^2$.
As explained below, $g^2$ is set in the $3d$ and $2d$ cases
by considering its relation to the string tension $\sqrt{\sigma}$,
while in the $4d$ case we adopt the value of $g^2(\omu^2)$ at the scale
$\omu$ in the MOM scheme.

In order to normalize the analytic expressions for $G(p^2)$ consistently with
the lattice data, we consider
two possibilities. In the first case, we take
\begin{equation}
F_1(p^2) \, =\, \frac{
1 - \sigma(\omu^2) }{p^2 \, \left[ 1 - \sigma(p^2) \right] } \; .
\label{eq:fnormal}
\end{equation}
Alternatively, as already discussed above in Section \ref{sec:4dtheory},
one can normalize $G(p^2)$ by adding a constant to $\sigma(p^2)$,
i.e.\ considering
\begin{equation}
F_2(p^2) \, =\,
\frac{1}{p^2 \, \left[ 1 - \sigma(p^2) + \sigma(\omu^2) \right] }\,.
\label{eq:fnormal2}
\end{equation}
Let us stress that, with the parameters fixed as above, these functions
are not fitting forms, but analytic predictions for $G(p^2)$ from previously
obtained (gluon-propagator) results. These will allow a good description
of the lattice data in the ultraviolet (UV) regime only. Nevertheless,
by treating $g^2$ as a free parameter in the above formulae and keeping
the remaining parameters fixed, one obtains
good-quality fits for the whole range of data in all cases. We indicate
the corresponding fitting forms by ${\widetilde F}_1(p^2)$ and
${\widetilde F}_2(p^2)$.


\subsection{The three-dimensional case}

\begin{figure}
\begin{center}
\includegraphics[width=.48\textwidth]{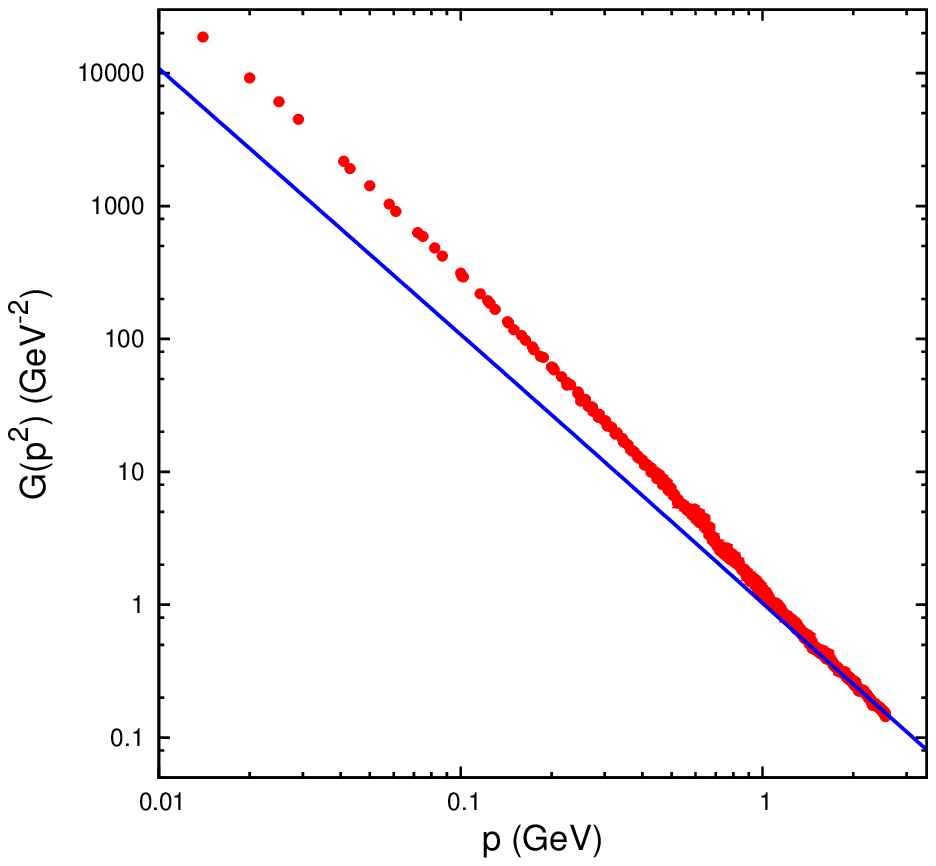}
\hskip 5mm
\includegraphics[width=.48\textwidth]{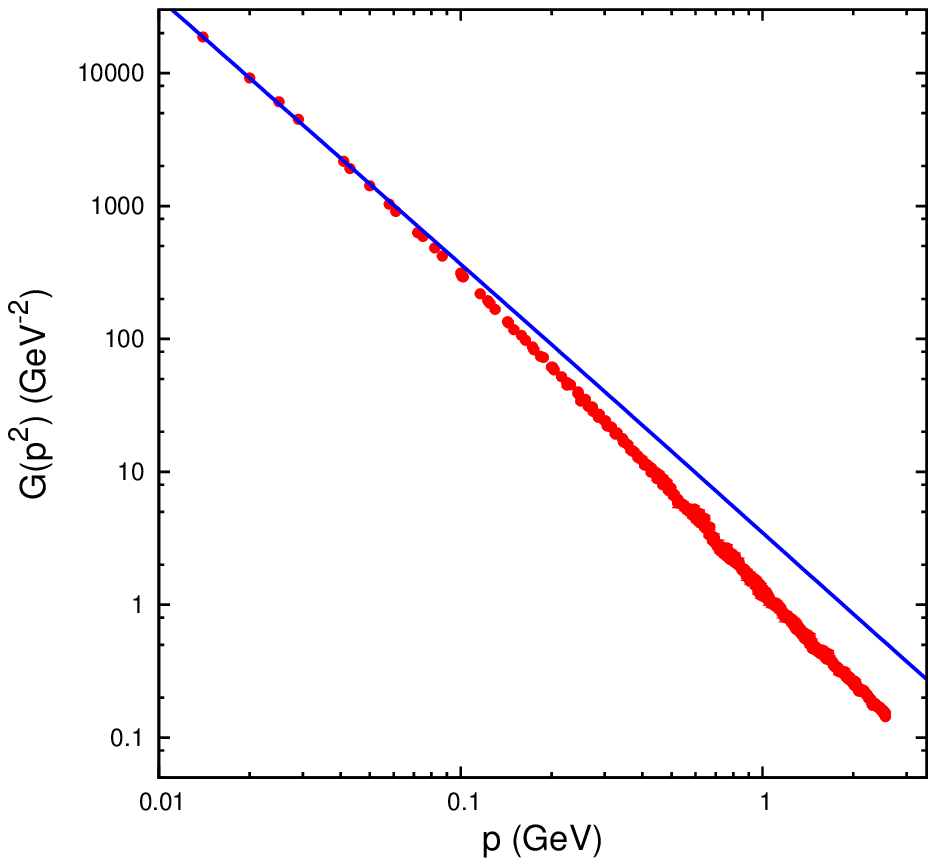}
\caption{Plot of the ghost propagator $G(p^2)$ as a function of the
magnitude of the (unimproved) momenta $p$ (both in physical units)
for the lattice volume $V =320^3$ at $\beta = 3.0$.
The data are (multiplicatively)
normalized to $1/\omu^2$ for $p=\omu=2.5 \,\text{GeV}$. We also
show the function $F_1(p^2)$ [see Eq.\ (\ref{eq:fnormal})]
(normalized in the same way) with the Gribov ghost form factor
$\sigma(p^2)$ given by the one-loop results
(\ref{eq:sigma3d2})--(\ref{eq:r3d}); the corresponding parameters are
reported in Eq.\ (\ref{eq:3d-param}) and we set $g^2 = 1.313 \, \text{GeV}$.
On the other hand, in the right plot, we fix the analytic form to match
the numerical result at $p = p_{min} = 14 \, \text{MeV}$, the
smallest nonzero (lattice) momentum for the pair $(V,\beta)$ considered,
yielding $3.38353 \, F_1(p^2)$.
Notice the logarithmic scale on both axes.
}
\label{fig:gren}
\end{center}
\end{figure}

\begin{figure}
\begin{center}
\includegraphics[width=.48\textwidth]{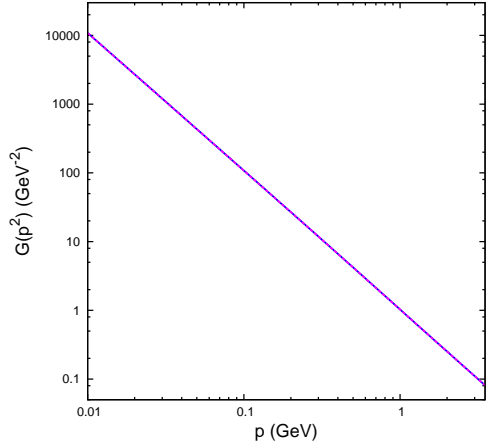}
\caption{Plot of $F_1(p^2)$ [see Eq.\ (\ref{eq:fnormal}), blue curve]
and $F_2(p^2)$ [see Eq.\ (\ref{eq:fnormal2}), magenta curve] as functions
of the momentum $p$ for the $3d$ case, with $\sigma(p^2)$ given
by $\sigma_{1L}(p^2)$ [see Eqs.\ (\ref{eq:sigma3d2})--(\ref{eq:r3d}) and
(\ref{eq:3d-param})] with $g^2 = 1.313 \, \text{GeV}$. For both curves we
consider $\omu=2.5 \,\text{GeV}$. Notice the logarithmic scale on both axes.
}
\label{fig:gcomparison}
\end{center}
\end{figure}

As discussed above, we now try
to describe the ghost-propagator data in $d=3$ by considering
the Gribov ghost form factor
$\sigma(p^2)$ given by $\sigma_{1L}(p^2)$ [see Eqs.\
(\ref{eq:sigma3d2})--(\ref{eq:r3d})].
We set the parameters $\alpha, \omega_1, a, b, v, w$ to the
values obtained in Refs.\ \cite{Cucchieri:2011ig,Cucchieri:2012gb}
by fitting the gluon propagator. In particular, we use the
values reported in Table XI of Ref.\ \cite{Cucchieri:2011ig} (from a
Monte Carlo analysis), i.e.\
\begin{eqnarray}
a=0.216(2)~\text{GeV}\, , \quad
b &=& 0.271(3)~\text{GeV}\, , \quad
\nu=0.215(5)~\text{GeV}^2\, , \quad
\nonumber \\[2mm]
w=0.580(6)~\text{GeV}^2\, , \quad
\alpha &=& -0.024(5)~\text{GeV}\, , \quad
\omega_1^2=0.046(4)~\text{GeV}^2 \; .
\label{eq:3d-param}
\end{eqnarray}
As for the bare coupling $g^2$, since in $3d$ it is a constant
(mass) parameter,\footnote{Let us recall that, in the general
$d$-dimensional case, we have that $g^2$ has mass dimension $4-d$.}
we use the SU(2) ratio $\sqrt{\sigma} / g^2 = 0.3351 (16)$
[see Eq.\ (7) of Ref.\ \cite{Lucini:2002wg}].
Then, with $\sqrt{\sigma} \approx 0.44 \, \text{GeV}$ we find
$g^2 \approx 1.313 \, \text{GeV}$.
The corresponding plot of $F_1(p^2)$ (see above)
is shown in Fig.\ \ref{fig:gren} (left plot).
Let us point out that for the momentum range spanned by the data the
functions $F_1(p^2)$ and $F_2(p^2)$ are numerically indistinguishable
(see Fig.\ \ref{fig:gcomparison}). Note that
$\sigma_{1L}(\omu^2) \approx 0.0252$ and that $\sigma_{1L}(p^2)$ takes
values\footnote{As shown in Ref.\ \cite{Cucchieri:2012cb}, from Eq.\
(\ref{eq:Gini}) one can write
\begin{equation}
\sigma(p^2) \, = \, \frac{N_c p_{\mu} p_{\nu}}{p^2} \,
       \int \frac{\d^d q}{(2\pi)^d}
       \, D(q^2) \, P_{\mu \nu}(q)
       \, \frac{1}{(p - q)^2}
\label{eq:sigmaini}
\end{equation}
and prove that $\partial \sigma(p^2) / \partial p^2 \,<\,0$ if the gluon
propagator $D(p^2)$ is positive in momentum space, i.e.\ $\sigma(p^2)$
---evaluated at one loop--- is monotonically decreasing as the momentum
$p$ increases.} in $[0.0247, 0.1014]$ when $p \in
[0.014, 2.553] \; \text{GeV}$.
Let us also mention that the one-loop expression $p^2\,F_1(p^2)$
does not change appreciably in the considered momentum range. (It goes
from about 1.0 in the UV to about 1.1 in the IR regimes.) Thus, the
momentum dependence of the analytic prediction $F_1(p^2)$ is almost
entirely due to the factor $1/p^2$.

\begin{figure}
\begin{center}
\includegraphics[width=.48\textwidth]{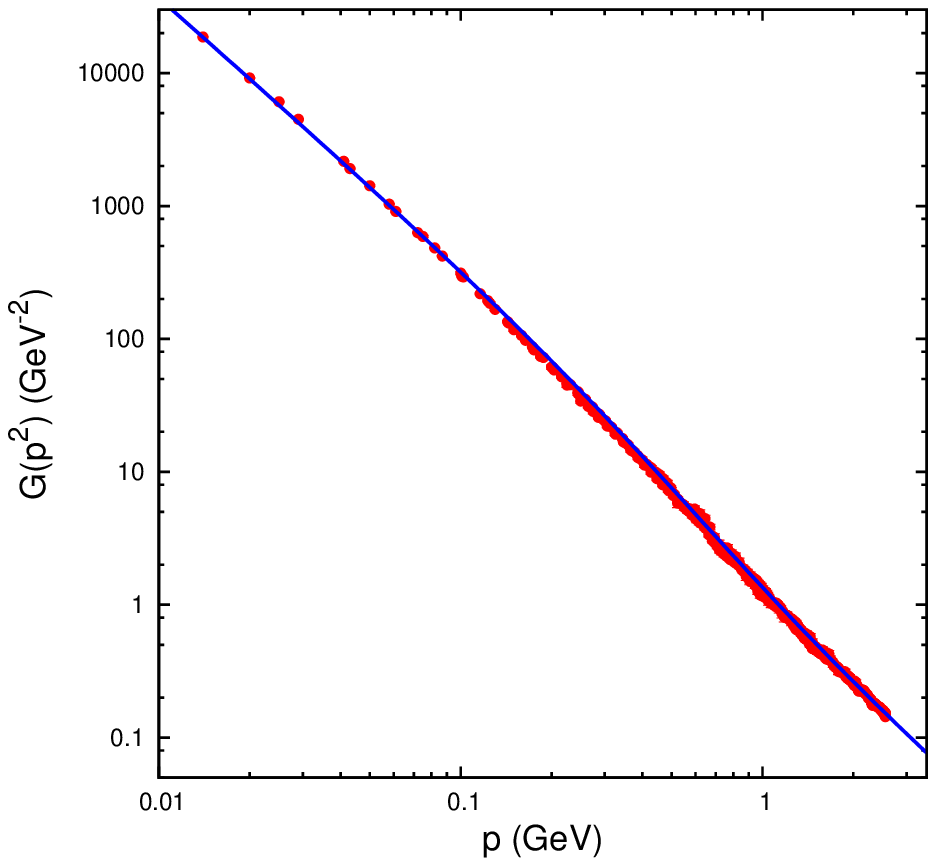}
\hskip 5mm
\includegraphics[width=.48\textwidth]{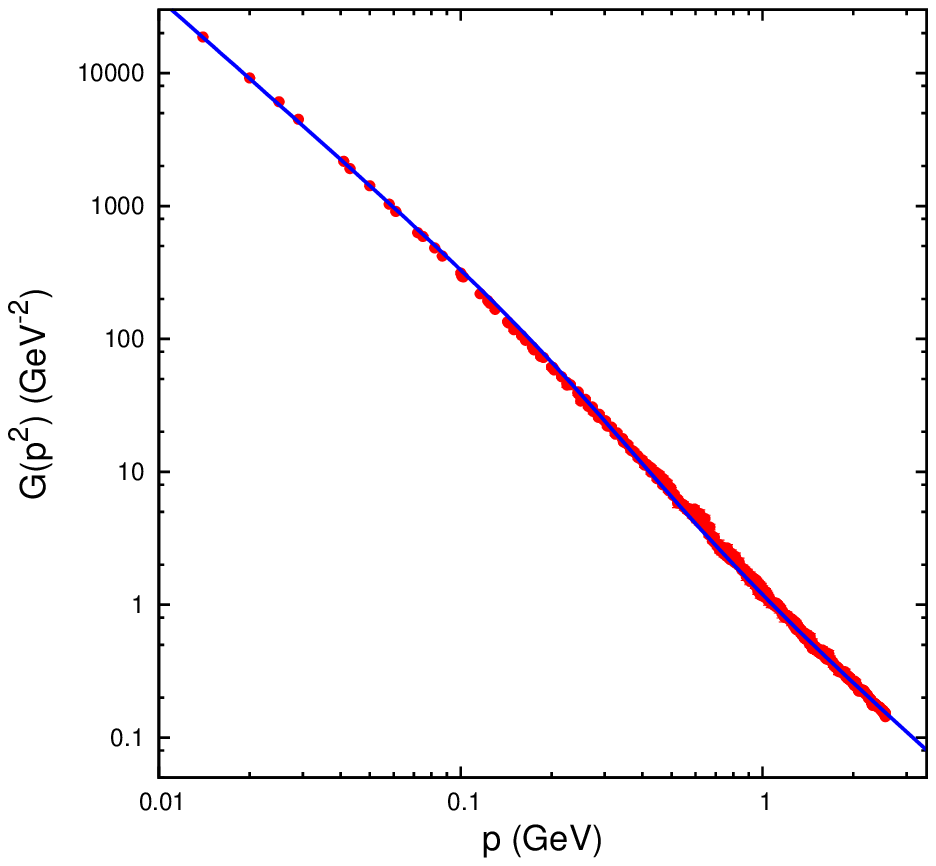}
\caption{Plot of the ghost propagator $G(p^2)$ as a function of the
magnitude of the (unimproved) momenta $p$ (both in physical units)
for the lattice volume $V =320^3$ at $\beta = 3.0$ together with the
fitting forms discussed in the text.
The data are (multiplicatively)
normalized to $1/\omu^2$ for $p=\omu=2.5 \,\text{GeV}$.
In the left plot we
show the function ${\widetilde F}_1(p^2)$
(normalized in the same way) with the Gribov ghost form factor
$\sigma(p^2)$ given by the one-loop results
(\ref{eq:sigma3d2})--(\ref{eq:r3d}); the corresponding parameters are
reported in Eq.\ (\ref{eq:3d-param}) and we use the fitted value
$10.0831 \, \text{GeV}$ for $g^2$.
On the other hand, in the right plot, we show the fitting function
$F_3(p^2)$ [see Eq.\ (\ref{eq:FG})] with the parameters given in Eqs.\
(\ref{eq:paramz})--(\ref{eq:params}). Notice the logarithmic scale
on both axes.
}
\label{fig:gfit}
\end{center}
\end{figure}

One can observe that, modulo a global factor, $F_1(p^2)$
has the expected leading UV and IR behaviors. Indeed, as shown
in Fig.\ \ref{fig:gren} (right plot), it agrees with the data in the IR limit
if we consider $3.38353 \, F_1(p^2)$. This implies that, in the deep IR limit,
$G(p^2)$ is enhanced by a finite multiplicative factor with
respect to the UV behavior. As mentioned above,
one can improve the description of the
ghost-propagator data in the whole momentum range by fitting the
values of $g^2$, instead of using a fixed value. In this case, we find
\begin{equation}
g^2 \, = \, 10.08 \pm 0.01 \; \text{GeV}
\label{eq:g23}
\end{equation}
with $\chi^2/dof \approx 4.5$ (with 480 data points). The corresponding
plot of ${\widetilde F}_1(p^2)$ is shown in Fig.\ \ref{fig:gfit} (left plot).
Let us stress that, with this fitted value for $g^2$, the analytic
prediction $p^2\,{\widetilde F}_1(p^2)$ varies from about 1.0 at
large momentum to
about 3.6 in the IR limit, a behavior that can be related to the
global rescaling shown in Fig.\ \ref{fig:gren}.
An even better fit of the data can be obtained with the fitting function
\cite{Cucchieri:2008fc}
\begin{equation}
F_3(p^2) \,=\, \frac{z}{p^2}
   \,\frac{t + p^2/s^2
    + \log\left(1 + p^2/s^2 \right)}{
              1 + p^2/s^2}
\label{eq:FG} \; ,
\end{equation}
inspired by Ref.\ \cite{Aguilar:2008xm}, which has $1/p^2$ leading IR
and UV behaviors. Indeed, with the fitting parameters set to
\begin{eqnarray}
z & = & 0.958 \pm 0.004 \label{eq:paramz} \\
t & = & 3.81 \pm 0.02 \label{eq:paramt} \\
s & = & 0.207 \pm 0.003 \; \text{GeV} \label{eq:params}
\end{eqnarray}
we find a $\chi^2/dof \approx 2.9$ (again with 480 data points). The
corresponding plot is shown in Fig.\ \ref{fig:gfit} (right plot).
Note that the value of the parameter $t$ is compatible with the multiplicative
constant obtained above when comparing the IR and UV behaviors of $F_1(p^2)$
(see Fig.\ \ref{fig:gren}).

\begin{figure}
\begin{center}
\includegraphics[width=.48\textwidth]{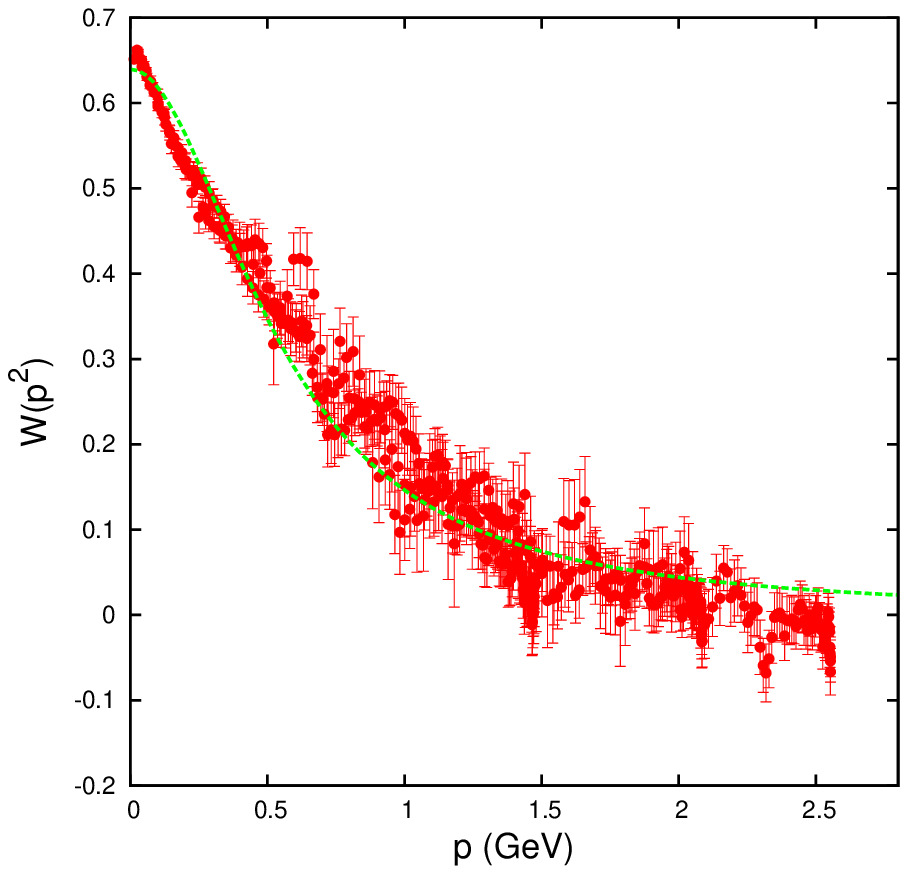}
\caption{Plot of the term $W(p^2)$ [see Eqs.\ (\ref{eq:G3dW}) and
(\ref{eq:w3d})] as a function of the (unimproved) momenta $p$
(in physical units) for the lattice volume $V =320^3$ at $\beta = 3.0$
(data points).
We also show (in red) the fitting function $\widetilde{W}(p^2)$
[see Eq.\ (\ref{eq:W})] with the fitting parameters reported in Eq.\
(\ref{eq:paramAB}).
}
\label{fig:sigmadif}
\end{center}
\end{figure}

One can try to estimate what is missing in the RGZ one-loop analysis for
$G(p^2)$, e.g.\ using the expression for $F_2(p^2)$ in Eq.\ (\ref{eq:fnormal2}).
More precisely, let us define a function $W(p^2)$ by
\begin{equation}
G(p^2) \, = \, \frac{1}{p^2 \, \left[1 - \sigma(p^2)
                     + \sigma(\omu^2) - W(p^2) \right]}
\label{eq:G3dW}
\end{equation}
and then use our numerical data for $G(p^2)$ [and the one-loop expression
for $\sigma(p^2)$] to get an estimate for $W(p^2)$.
To this end, we carried out a Monte Carlo analysis (with 10000 samples) of
the quantity
\begin{equation}
W(p^2) \,=\,\left[\,1 \,- \, \frac{1}{p^2 \, G(p^2)}\,\right]
                 \,-\,\sigma_{1L}(p^2) \,+\, \sigma_{1L}(\omu^2) \; ,
\label{eq:w3d}
\end{equation}
where $G(p^2)$ represents the numerical (multiplicatively normalized) ghost
propagator result at a given momentum $p$ and $\sigma_{1L}(p^2)$ is the
one-loop estimate (\ref{eq:sigma3d2})--(\ref{eq:r3d}) with
the parameters given in Eq.\ (\ref{eq:3d-param}) and the value of
$g^2$ set to $1.313 \, \text{GeV}$.
The corresponding plot is shown
in Fig.\ \ref{fig:sigmadif}. The estimated error for $W(p^2)$
includes the error in the data points for $G(p^2)$ and the errors in
the parameters.
Note that $W(p^2)$ goes from approximately zero in the UV regime to about
0.7 in the IR limit, which is consistent with the small variation of
$p^2\,F_1(p^2)$ discussed above.

One can parametrize the function $W(p^2)$ reasonably well
by using the simple expression
\begin{equation}
\widetilde{W}(p^2) \, = \, \frac{A}{1 + B \, p^2}
\label{eq:W}
\end{equation}
with
\begin{equation}
A \, \approx  \, 0.64 \label{eq:paramAB} \;\; ,
\qquad \;
B \, \approx \, 3.4  \; \text{GeV}^{-2} \; ,
\end{equation}
yielding a $\chi^2/dof$ of 2.7 (with 480 data points). The
corresponding plot (red curve) is also shown in Fig.\ \ref{fig:sigmadif}.


\subsection{The four-dimensional case}

\begin{figure}
\begin{center}
\includegraphics[width=.48\textwidth]{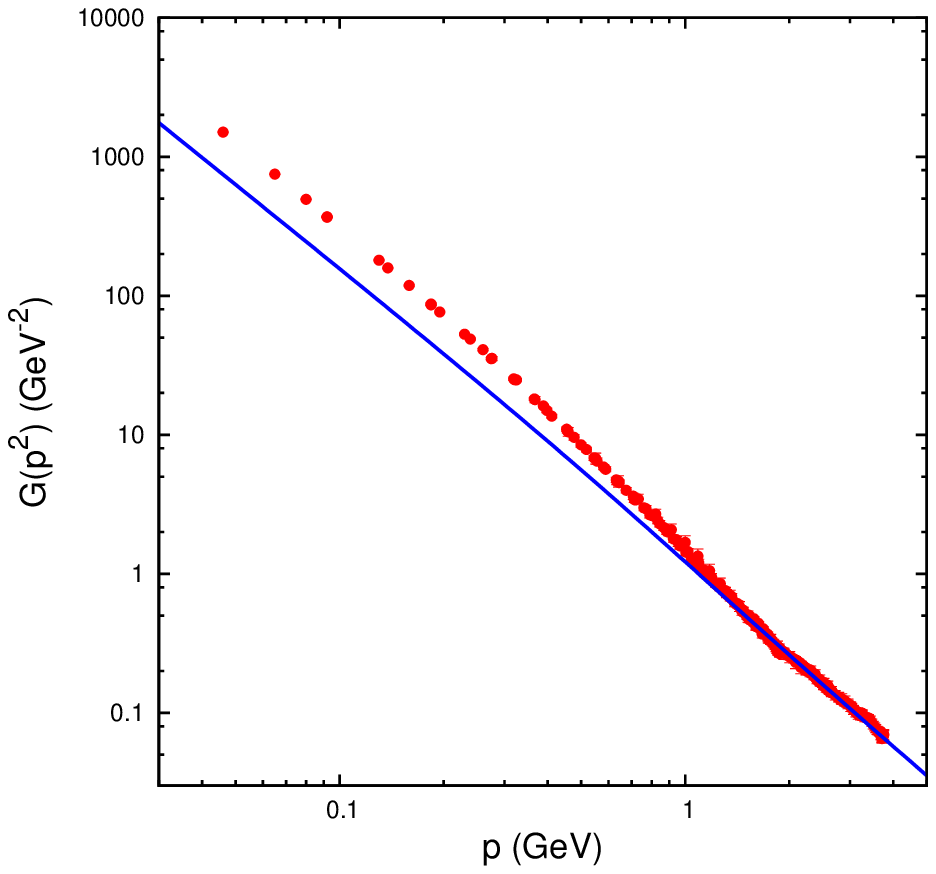}
\hskip 5mm
\includegraphics[width=.48\textwidth]{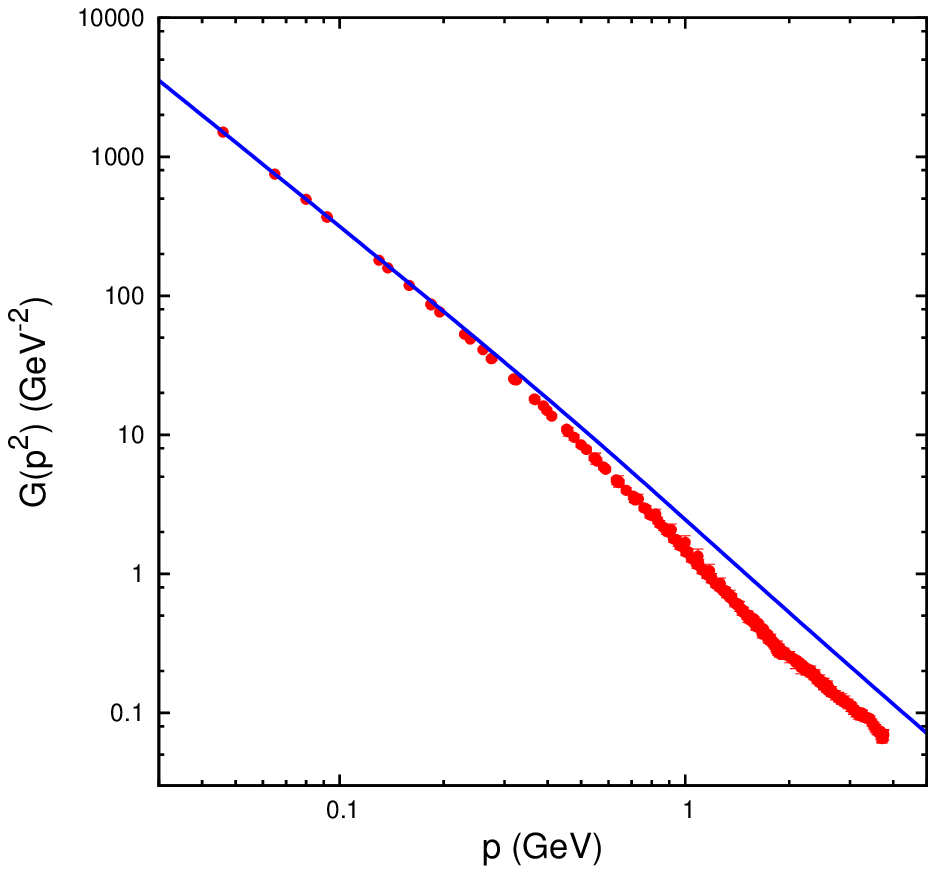}
\caption{
Plot of the ghost propagator $G(p^2)$ as a function of the
magnitude of the (unimproved) momenta $p$ (both in physical units)
for the lattice volume $V =128^4$ at $\beta = 2.2$.
The data are (multiplicatively)
normalized to $1/\omu^2$ for $p=\omu=2.5 \,\text{GeV}$. We also
show the function $F_2(p^2)$ [see Eq.\ (\ref{eq:fnormal2})]
(normalized in the same way) with the Gribov ghost form factor
$\sigma(p^2)$ given by the one-loop results (\ref{eq:sigma4d2})--(\ref{eq:a3});
the corresponding parameters are
reported in Eq.\ (\ref{eq:param4d}) and we set $g^2 = 7.794$.
On the other hand, in the right plot, we fix the analytic form to match
the numerical result at $p = p_{min} = 46 \,\text{MeV}$, the smallest
nonzero (lattice) momentum for the pair $(V,\beta)$ considered,
yielding $2.01654 \, F_2(p^2)$.
Notice the logarithmic scale on both axes.
}
\label{fig:gren4d}
\end{center}
\end{figure}

For the $4d$ case we repeat the same type of analysis carried out
in the previous section for the $3d$ case. In particular,
as explained in Section \ref{sec:4dtheory} above, we consider the function
$F_2(p^2)$ in Eq.\ (\ref{eq:fnormal2}) with the one-loop expression for
$\sigma(p^2)$
given by $\sigma^{\overline{\mbox{\tiny{MS}}}}_{1L}(p^2)$ [see Eqs.\
(\ref{eq:sigma4d2})--(\ref{eq:a3})] and $\omu = 2.5 \,\text{GeV}$.
Again, by using the (gluon-propagator) results presented in Refs.\
\cite{Cucchieri:2011ig, Cucchieri:2012gb}, the parameters $a, b, \nu, w$
are set to the values reported in Table IV of Ref.\
\cite{Cucchieri:2011ig} and obtained using a Monte Carlo analysis,
i.e.\
\begin{equation}
\label{eq:param4d}
a=0.392(2)\, , \quad
b=1.32(5)\, , \quad
\nu=0.29(2)~\text{GeV}^2\, , \quad
w=0.66(1)~\text{GeV}^2 \; .
\end{equation}
Here we can estimate the value of $g^2$, at a given scale $\omu$ and
in the MOM scheme, by considering the one-loop result
\begin{equation}
\label{4D13}
g^2(\omu) \, = \, \frac{1}{\beta_0 \,
   \ln\left( \frac{\omu^2}{\Lambda_{\mbox{\tiny{MOM}}}^2} \right)}
\end{equation}
with $\beta_0 = 11 N_c / (48 \pi^2)$, which is valid for any SU($N_c$)
gauge group. Then, the value of $\Lambda_{\mbox{\tiny{MOM}}}$ can be
obtained by considering the relation \cite{Celmaster:1979km}
\begin{equation}
g^2 \, = \, \overline{g}^2 \, \left( \, 1 \, + \, \frac{169 N_c}{36}
        \, \overline{g}^2 \, + \, \ldots \, \right)
\end{equation}
between the MOM-scheme coupling $g^2$ and the $\MSbar$ coupling
$\overline g^2$. This implies (see for example \cite{Boucaud:2008gn})
$\Lambda_{\mbox{\tiny{MOM}}} = \lms \, e^{169/264}$, which is valid
for any value of $N_c$ and with $N_f=0$, where $N_f$ is the number of
quark flavors. For the SU(2) case, i.e.\ for $N_c = 2$, one can use the
estimate $\lms \approx 0.752\, \sqrt{\sigma}$ (see Ref.\
\cite{Lucini:2008vi}), where $\sqrt{\sigma}$ is the string tension.
Then, after setting $\sqrt{\sigma}\approx 0.44~\text{GeV}$ we find
$\lms \approx 331\,\text{MeV}$ and $\Lambda_{\mbox{\tiny{MOM}}}
\approx 628 \, \text{MeV}$. For the subtraction point $\omu = 2.5\,
\text{GeV}$, used here, this gives for the effective MOM coupling a
value of $g^2(\omu) \approx 7.794$, which yields $\alpha_s(\omu) =
g^2(\omu) / (4 \pi) \approx 0.6202$. The corresponding plot of $F_2(p^2)$
is shown in Fig.\ \ref{fig:gren4d} (left plot).
Also in this case, the functions $F_1(p^2)$ in Eq.\ (\ref{eq:fnormal})
and $F_2(p^2)$ in Eq.\ (\ref{eq:fnormal2}) are numerically
indistinguishable. Note that $\sigma^{\overline{\mbox{\tiny{MS}}}}_{1L}(\omu^2)
\approx 0.1419$ and that $\sigma^{\overline{\mbox{\tiny{MS}}}}_{1L}(p^2)$
takes values in $[0.06502, 0.5081]$ when $p \in [0.046, 3.752] \;
\text{GeV}$, which is the momentum interval for which we have numerically
evaluated the ghost propagator $G(p^2)$.
Here, contrary to the $3d$ case, the one-loop expression $p^2\,F_2(p^2)$
is not flat with the momentum $p$, i.e.\ it
changes from about 0.9 in the UV to about 1.6 in the IR regimes.

\begin{figure}
\begin{center}
\includegraphics[width=.48\textwidth]{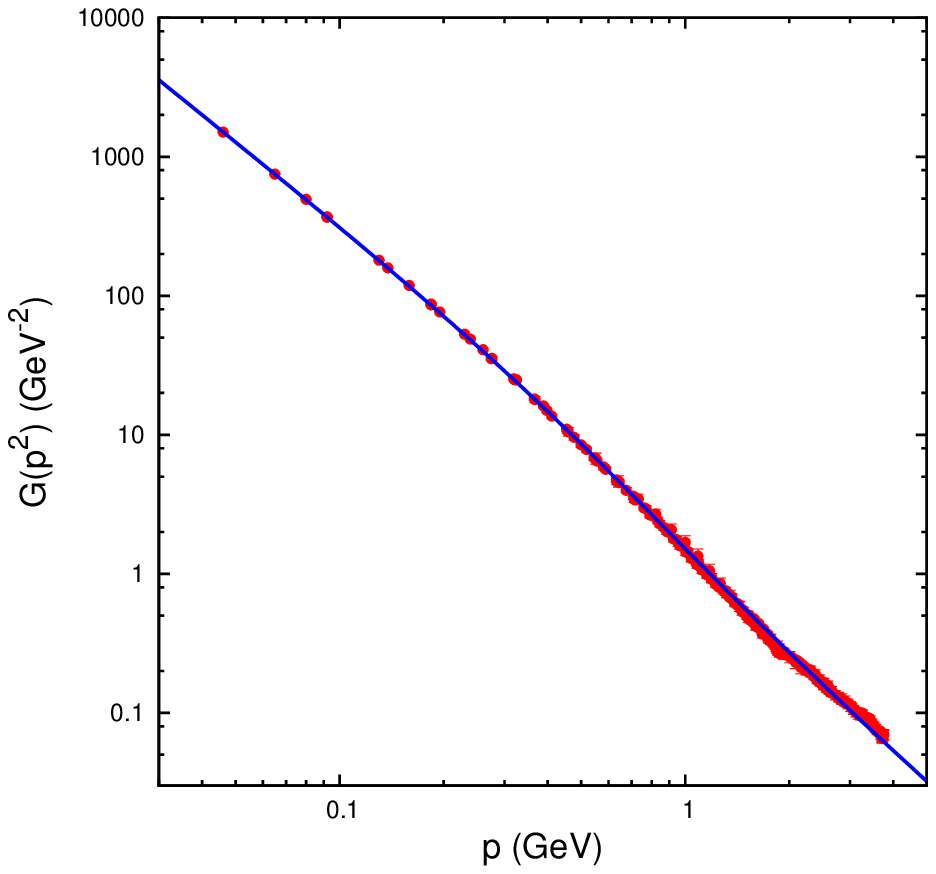}
\hskip 5mm
\includegraphics[width=.48\textwidth]{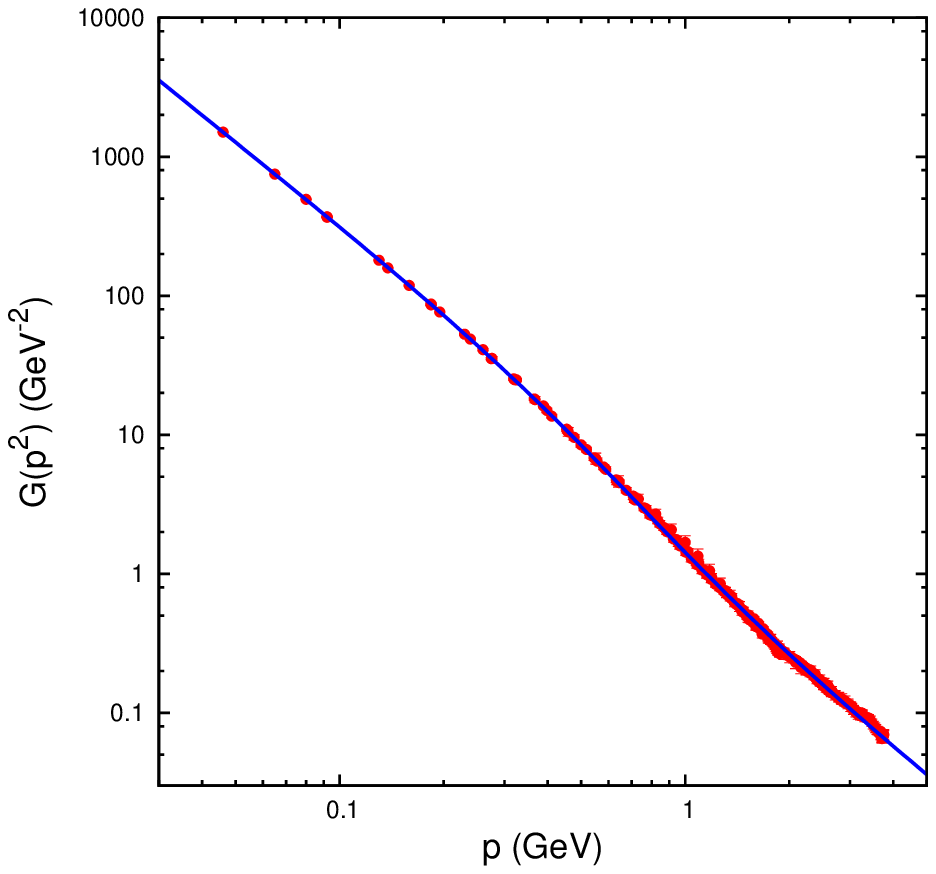}
\caption{Plot of the ghost propagator $G(p^2)$ as a function of the
magnitude of the (unimproved) momenta $p$ (both in physical units)
for the lattice volume $V =128^4$ at $\beta = 2.2$ together with the
fitting forms discussed in the text.
The data are (multiplicatively)
normalized to $1/\omu^2$ for $p=\omu=2.5 \,\text{GeV}$.
In the left plot we
show the function ${\widetilde F}_2(p^2)$
(normalized in the same way) with the Gribov ghost form factor
$\sigma(p^2)$ given by the one-loop results
(\ref{eq:sigma4d2})--(\ref{eq:a3}); the corresponding parameters are
reported in Eq.\ (\ref{eq:param4d}) and we use the fitted value
$14.6165$ for $g^2$.
On the other hand, in the right plot, we show the fitting function
$F_3(p^2)$ [see Eq.\ (\ref{eq:FG})] with the parameters given in Eqs.\
(\ref{eq:paramz4d})--(\ref{eq:params4d}). Notice the logarithmic scale
on both axes.
}
\label{fig:gfit4d}
\end{center}
\end{figure}

\begin{figure}
\begin{center}
\includegraphics[width=.48\textwidth]{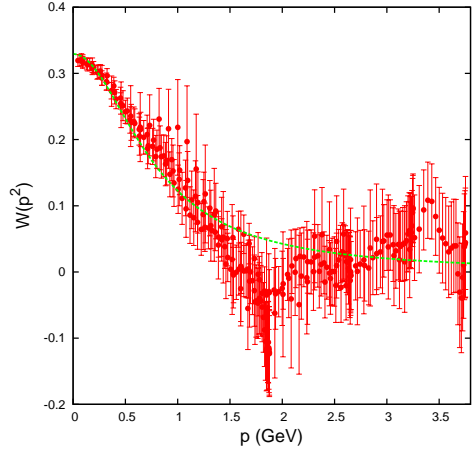}
\caption{Plot of the term $W(p^2)$ [see Eqs.\ (\ref{eq:G3dW}) and
(\ref{eq:w3d})] as a function of the (unimproved) momenta $p$
(in physical units) for the lattice volume $V =128^4$ at $\beta = 2.2$
(data points).
We also show (in red) the fitting function $\widetilde{W}(p^2)$
[see Eq.\ (\ref{eq:W})] with the fitting parameters reported in Eq.\
(\ref{eq:paramAB4d}).
}
\label{fig:sigmadif4d}
\end{center}
\end{figure}

Qualitatively, the situation in the $4d$ case is very similar to what we have
seen above in the $3d$ case. In particular, one can obtain a good description
of the numerical data in the IR limit by rescaling the analytic prediction
$F_2(p^2)$ by the factor 2.01654 (see right plot of Fig.\ \ref{fig:gren4d})
and a good description of all the data by fitting the values of $g^2$. Indeed,
with
\begin{equation}
g^2 \, = \, 14.62 \pm 0.01
\label{eq:g24}
\end{equation}
we obtain a $\chi^2/dof \approx 1.7$ (with 256 data points). The corresponding
plot of ${\widetilde F}_2(p^2)$ is shown in Fig.\ \ref{fig:gfit4d} (left plot).
An even better fit (see right plot in Fig.\ \ref{fig:gfit4d}) is obtained
with the fitting function (\ref{eq:FG}) and the parameters set to
\begin{eqnarray}
z & = & 0.859 \pm 0.006 \label{eq:paramz4d} \\
t & = & 3.73 \pm 0.02 \label{eq:paramt4d} \\
s & = & 0.407 \pm 0.005 \; \text{GeV} \;, \label{eq:params4d}
\end{eqnarray}
which yields a $\chi^2/dof \approx 0.75$ (again with 256 data points).
Here the value of $t$ can be related to the global rescaling shown on
the right in Fig.\ \ref{fig:gren4d} (i.e.\ approximately a factor 2)
and to the above mentioned change in
$p^2\,F_2(p^2)$, yielding a factor $2\times 1.6/0.9\approx 3.6\,$.
The parameter $t$ can also be related to the variation of
$p^2\,{\widetilde F}_2(p^2)$
from about 0.9 at large momentum to about 3.2 in the IR limit, yielding
a factor $3.2/0.9\approx 3.6\,$.

Finally, in Fig.\ \ref{fig:sigmadif4d} we present the numerical estimate
---using a Monte Carlo analysis with 10000 samples--- for the quantity
$W(p^2)$, defined in Eq.\ (\ref{eq:w3d}) and using the $4d$ one-loop
expression for $\sigma(p^2)$, as well as the fitting function
$\widetilde{W}(p^2)$, defined in Eq.\ (\ref{eq:W}). With the values
\begin{equation}
A \, \approx \, 0.33  \label{eq:paramAB4d} \;\, ,
\qquad
B \, \approx \, 1.7  \; \text{GeV}^{-2} \;
\end{equation}
for the parameters we find a $\chi^2/dof$ of 0.97 (with 256 data
points). It is also interesting to note that, in this case, the magnitude
of what is missing in the one-loop calculation of $\sigma(p^2)$ is about
$50\%$ smaller than the corresponding outcome obtained in the $3d$ case.
This is expected since, as mentioned above, there is a larger change in
$p^2 F_2(p^2)$ over the momentum range in the $4d$ case.


\begin{figure}
\begin{center}
\includegraphics[width=.48\textwidth]{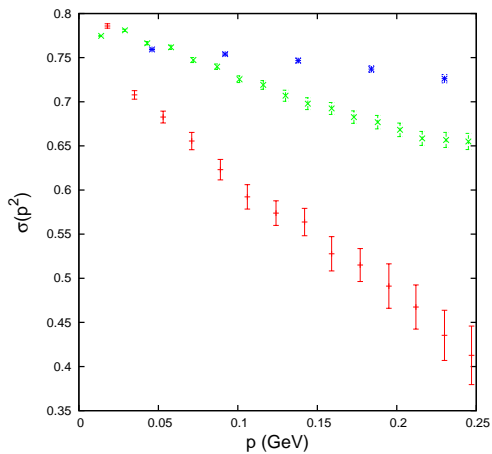}
\caption{Plot of the Gribov ghost form factor $\sigma(p^2)$ [see Eq.\
(\ref{eq:sigmabis})] for the lattice volumes $V = 320^2$ at $\beta = 10.0$
(symbol $+$ in red), $V = 320^3$ at $\beta =3.0$ (symbol $\times$ in
green) and $V = 128^4$ at $\beta = 2.2$ (symbol $*$ in blue),
as a function of the unimproved momenta [see Eq.\ (\ref{eq:punimpr})],
for momenta with only one component different from zero.
All momenta are in physical units and we show the data only in the
IR limit, i.e.\ for $p \leq 0.25~\text{GeV}$. Error bars have been
estimated using propagation of errors. One clearly sees that in
the $3d$ and $4d$ cases $\sigma(p^2)$ becomes almost
constant at small momenta, with a value $\sigma(0)< 1$, implying a
free-like behavior for the ghost propagator in the IR limit. On the
contrary, in the $2d$ case $\sigma(p^2)$ is still
clearly increasing for momenta of the order of $20~\text{MeV}$.
}
\label{fig:sigma-comp}
\end{center}
\end{figure}

\subsection{The two-dimensional case}

Finally, we consider data for the $2d$ case. As already
stressed in the Introduction, in this case the ghost propagator
is IR-enhanced (see also Fig.\ \ref{fig:sigma-comp}). Thus, the analysis
of the numerical data will be done following the same ideas presented
in the two subsections above, but with a different fitting function
instead of $F_3(p^2)$ in Eq.\ (\ref{eq:FG}).
Nevertheless, as a first step, we consider again the one-loop
result $F_1(p^2)$ [see Eq.\ (\ref{eq:fnormal})] with $\sigma(p^2)$
given by $\sigma_{2d}(p^2)$
defined in Eqs.\ (\ref{eq:sigma2d})--(\ref{eq:imaginaryw}) and with
$\omu = 2.5 \,\text{GeV}$. At the same time, the parameters $a, b, \nu,
w, c$ and $\eta$ are set
considering the outcomes presented in Refs.\
\cite{Cucchieri:2011ig,Cucchieri:2012gb}.
In particular, we used the values
reported in Table XIV of Ref.\ \cite{Cucchieri:2011ig} and obtained
using a Monte Carlo analysis, i.e.\
\begin{eqnarray}
a=0.0550(5)~\text{GeV}^2\, , \quad
b &=& -0.049(7)~\text{GeV}^2\, , \quad
\nu=0.145(8)~\text{GeV}^2\, , \quad \nonumber \\[2mm]
w=0.15(1)~\text{GeV}^2\, , \quad
c &=& 0.07(1)~\text{GeV}^{2-\eta}\, , \quad
\eta=0.91(5) \; .
\label{eq:2d-param}
\end{eqnarray}
Note that the bound $\eta < 2$ (see Sec.\ \ref{sec:2dtheory}) is respected
by the fitted value. As for the he coupling constant $g^2$, its value can
be estimated by employing the analytic evaluation of the string tension
$\sqrt{\sigma}$. Indeed, for two-dimensional SU(2) lattice gauge theory in
the infinite volume limit, one has \cite{Dosch:1978jt}
\begin{equation}
\sigma_{latt} \, = \, - \ln\left[ \frac{I_2(\beta)}{I_1(\beta)} \right] \; ,
\end{equation}
where $I_n(\beta)$ is the modified Bessel function \cite{GR}. For large $\beta$
values (in our case we have $\beta = 10$), this yields
$\sigma_{latt} \approx 3/ (2 \, \beta)$.
Then, using the relation $\beta = 2 N_c / (g^2 a^{4-d})$,
where $a$ is the lattice spacing and which is valid for the SU($N_c$)
gauge group in $d$ dimensions, we find in the $2d$ case
\begin{equation}
g^2 \, \approx \, \frac{4\, N_c\, \sigma_{latt}}{3\, a^2} \; .
\end{equation}
For $N_c = 2$ and using the continuum value $\sqrt{\sigma_{latt}}/a \approx
0.44~\text{GeV}$ we obtain $g^2 \approx 0.516~\text{GeV}^2$.

One can check that, in the $2d$ case, the functions $F_1(p^2)$
in Eq.\ (\ref{eq:fnormal}) and $F_2(p^2)$ in Eq.\
(\ref{eq:fnormal2}) are again numerically indistinguishable. Note that
$\sigma_{2d}(\omu^2) \approx 0.00179$ and $\sigma_{2d}(p^2)$ takes
values in $[0.00173, 0.0334]$ when $p \in [0.018, 2.553] \;
\text{GeV}$, which is the momentum interval for which we have numerically
evaluated the ghost propagator $G(p^2)$.
Thus, also in this case, the one-loop expression $p^2\,F_1(p^2)$
does not change appreciably in the considered momentum range. (It goes
from 1.00 in the UV to 1.03 in the IR regimes.) As a consequence, the
momentum dependence of the analytic prediction is entirely due to the factor
$1/p^2$ and in this case we should {\em not} expect a good description of
the data in the IR region.

\begin{figure}
\begin{center}
\includegraphics[width=.48\textwidth]{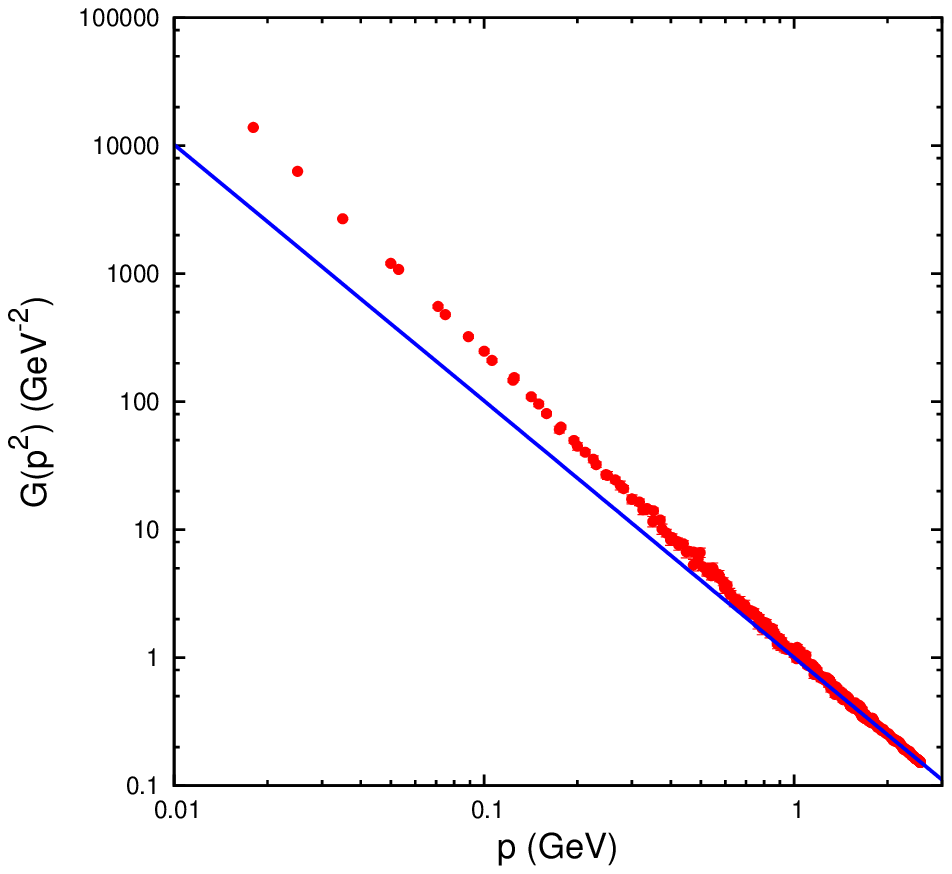}
\hskip 5mm
\includegraphics[width=.48\textwidth]{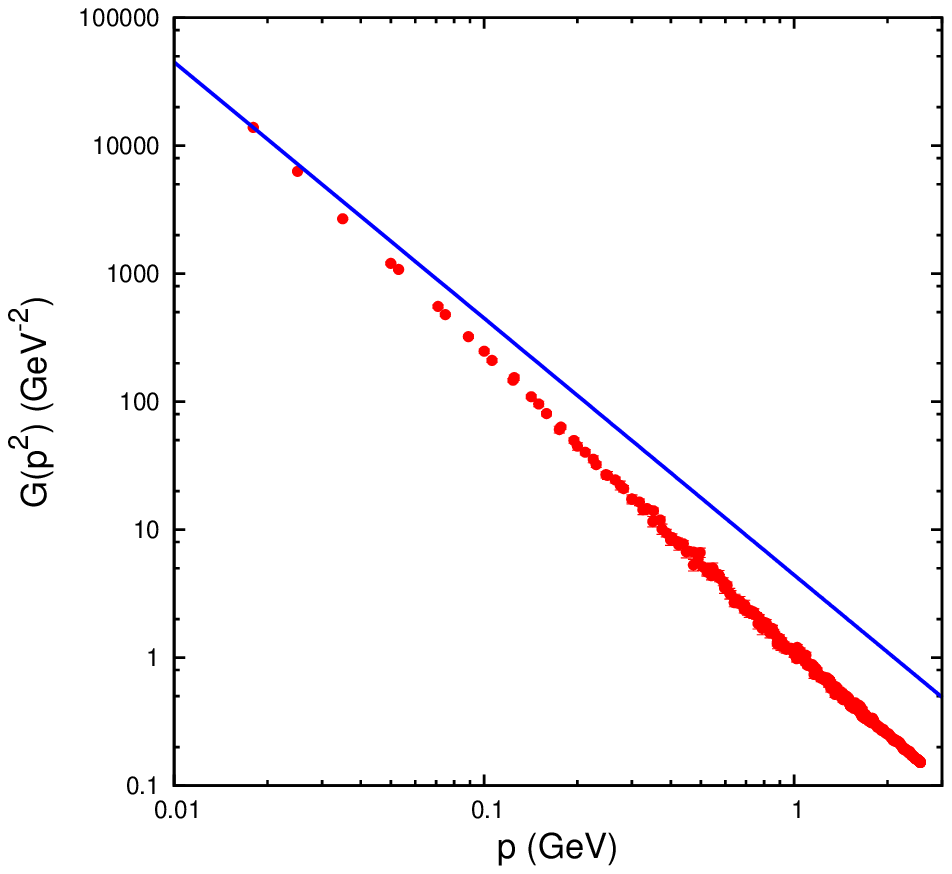}
\caption{Plot of the ghost propagator $G(p^2)$ as a function of the
magnitude of the (unimproved) momenta $p$ (both in physical units)
for the lattice volume $V =320^2$ at $\beta = 10.0$.
The data are (multiplicatively)
normalized to $1/\omu^2$ for $p=\omu=2.5 \,\text{GeV}$. We also
show the function $F_1(p^2)$ [see Eq.\ (\ref{eq:fnormal})]
(normalized in the same way) with the Gribov ghost form factor
$\sigma(p^2)$ given by the one-loop results
(\ref{eq:sigma2d})--(\ref{eq:imaginaryw}); the corresponding parameters are
reported in Eq.\ (\ref{eq:2d-param}) and we set $g^2 = 0.516 \, \text{GeV}^2$.
On the other hand, in the right plot, we fix the analytic form to match
the numerical result at $p = p_{min} = 18 \, \text{MeV}$, the
smallest nonzero (lattice) momentum for the pair $(V,\beta)$ considered,
yielding $4.41862 \, F_1(p^2)$.
Notice the logarithmic scale on both axes.
}
\label{fig:gren2d}
\end{center}
\end{figure}

\begin{figure}
\begin{center}
\includegraphics[width=.48\textwidth]{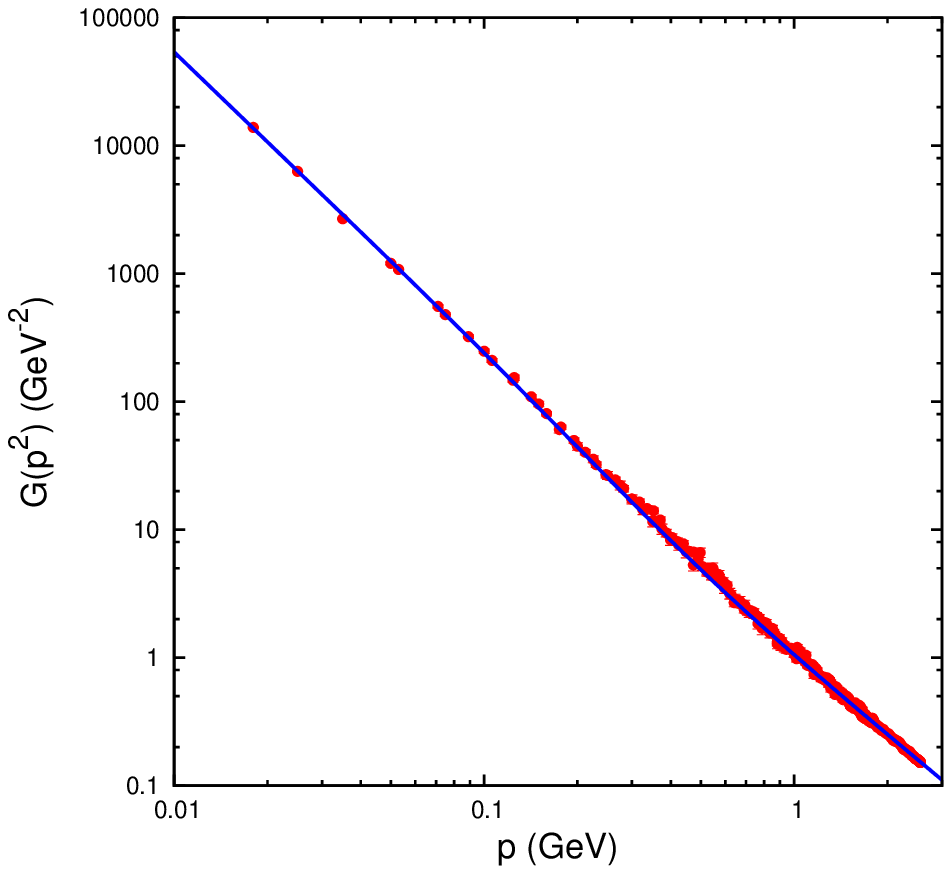}
\hskip 5mm
\includegraphics[width=.48\textwidth]{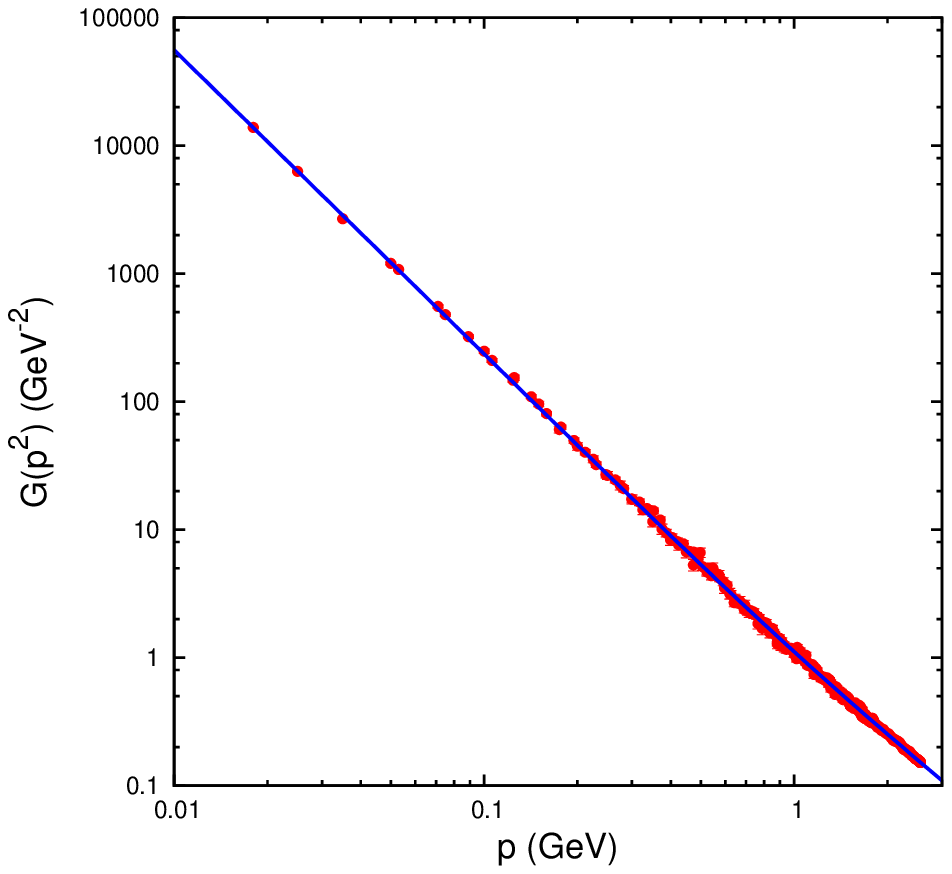}
\caption{Plot of the ghost propagator $G(p^2)$ as a function of the
magnitude of the (unimproved) momenta $p$ (both in physical units)
for the lattice volume $V =320^2$ at $\beta = 10.0$ together with the
fitting forms discussed in the text.
The data are (multiplicatively)
normalized to $1/\omu^2$ for $p=\omu=2.5 \,\text{GeV}$.
In the left plot we
show the function ${\widetilde F}_1(p^2)$
(normalized in the same way) with the Gribov ghost form factor
$\sigma(p^2)$ given by the one-loop results
(\ref{eq:sigma2d})--(\ref{eq:imaginaryw}); the corresponding parameters are
reported in Eq.\ (\ref{eq:2d-param}) and we use the fitted value
$13.4556 \, \text{GeV}^2$ for $g^2$.
On the other hand, in the right plot, we show the fitting function
$F_{2d}(p^2)$ [see Eq.\ (\ref{eq:FG2d})] with the parameters given in Eqs.\
(\ref{eq:paramz2d})--(\ref{eq:params2d}). Notice the logarithmic scale
on both axes.
}
\label{fig:gfit2d}
\end{center}
\end{figure}

As in the $3d$ and $4d$ cases, the analytic prediction gives a good
description of the data in the
UV limit (see left plot in Fig.\ \ref{fig:gren2d}). However, since the
value obtained for $g^2$ is smaller than the critical value $g^2_c$ ---i.e.\
the one-loop result is free-like at small momenta, while the numerical data
are IR-enhanced--- in the $2d$ case one cannot indeed describe well
the IR data by a simple global rescaling of the function
$F_1(p^2)$ (see right plot in Fig.\ \ref{fig:gren2d}). On the
other hand, by fitting $g^2$ ---i.e.\ considering the function
${\widetilde F}_1(p^2)$--- one finds that the value $g^2 =
13.46(2) ~\text{GeV}^2$ allows a good description of the lattice data (see
left plot in Fig.\ \ref{fig:gfit2d}) with $\chi^2/dof \approx 1.6$
and 320 data points. Let us stress that for $d=2$ choosing
the fitted value for $g^2$ over the fixed theoretical one has a
dramatic effect on the behavior of $p^2\,{\widetilde F}_1(p^2)$.
Indeed, this quantity
goes from about 1.0 at the largest momenta to about 9.0 in the IR limit.

Also, a slightly better fit can be obtained with the
function\footnote{This fitting function is inspired by the
one considered in Ref.\ \cite{Cucchieri:2008fc} for the $2d$
case, but with one less parameter. We have checked that the function employed
in Ref.\ \cite{Cucchieri:2008fc} allows only a modest improvement in
the description of the data when compared to the simpler fitting function
$F_{2d}(p^2)$ considered here.}
\begin{equation}
F_{2d}(p^2) \,=\, \frac{z}{p^2}
   \,\left( \frac{1 + p^2/s^2}{p^2/s^2} \right)^t
\label{eq:FG2d} \; ,
\end{equation}
Indeed, with the fitting parameters set to
\begin{eqnarray}
z & = & 0.963 \pm 0.002 \label{eq:paramz2d} \\
t & = & 0.188 \pm 0.002 \label{eq:paramt2d} \\
s & = & 1.08 \pm 0.04 \; \text{GeV} \label{eq:params2d}
\end{eqnarray}
we find $\chi^2/dof \approx 1.2$ (again with 320 data points). The
corresponding plot is shown in Fig.\ \ref{fig:gfit2d} (right plot).
Note that the factor $(s^2/p_{min}^2)^t\approx 4.7$ is compatible with
the multiplicative constant obtained above when comparing the IR and UV
behaviors of $F_1(p^2)$ (see right plot in Fig.\ \ref{fig:gren2d}).

\begin{figure}
\begin{center}
\includegraphics[width=.48\textwidth]{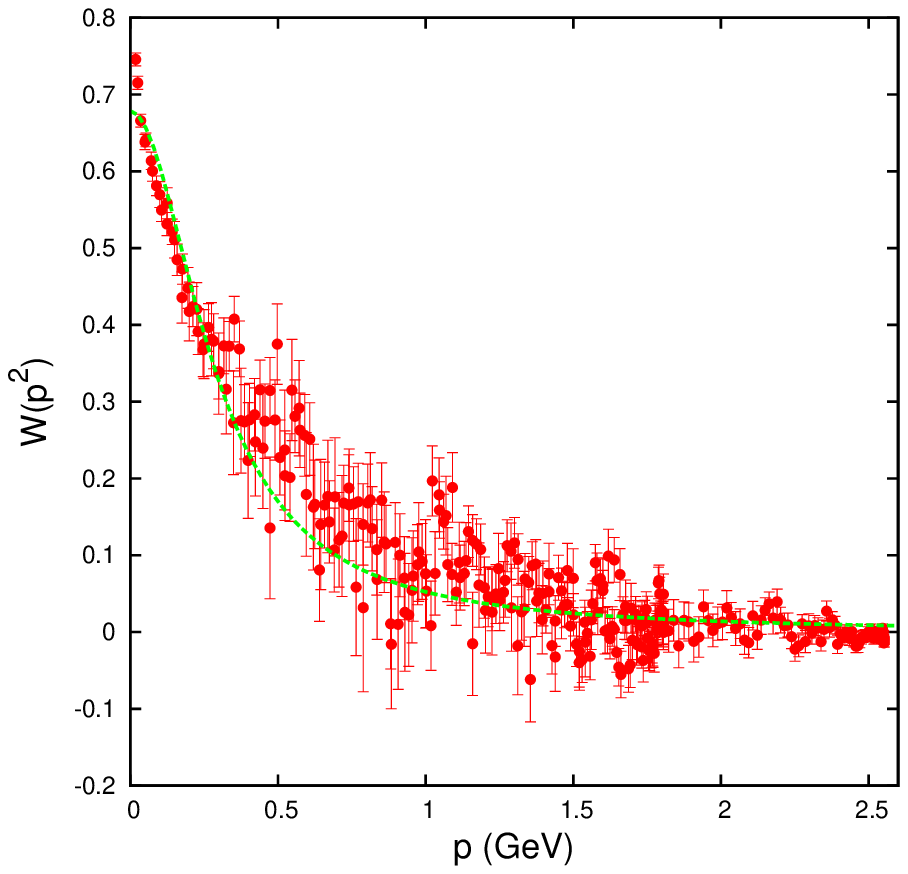}
\caption{Plot of the term $W(p^2)$ [see Eqs.\ (\ref{eq:G3dW}) and
(\ref{eq:w3d})] as a function of the (unimproved) momenta $p$
(in physical units) for the lattice volume $V =320^2$ at $\beta = 10.0$
(data points).
We also show (in red) the fitting function $\widetilde{W}(p^2)$
[see Eq.\ (\ref{eq:W})] with the fitting parameters reported in Eq.\
(\ref{eq:paramAB2d}).
}
\label{fig:sigmadif2d}
\end{center}
\end{figure}

As in the $3d$ and $4d$ cases, one can also estimate what is missing
in the one-loop analysis, i.e.\ we can evaluate $W(p^2)$ [see Eqs.\
(\ref{eq:G3dW}) and (\ref{eq:w3d}), using a Monte Carlo analysis with
10000 samples] as a function of the (unimproved) momenta $p$. The
corresponding data (see Fig.\ \ref{fig:sigmadif2d}) can be reasonably
described by the fitting function $\widetilde{W}(p^2)$ [see Eq.\
(\ref{eq:W})] with the fitting parameters
\begin{equation}
A \, \approx \, 0.68 \label{eq:paramAB2d} \; \;,
\qquad
B \, \approx \, 12.0  \; \text{GeV}^{-2} \; ,
\end{equation}
which yields a $\chi^2/dof$ of 2.5 (with 320 data points).
It is also interesting to note that, in this case, as for $d=3$,
the magnitude of what is missing in the one-loop calculation of
$\sigma(p^2)$ is quite large, since $p^2\,F_1(p^2)$ is essentially constant.

\section{Conclusions}
\label{sec:conclusions}

We have presented the final step of our analysis of large-lattice
Landau-gauge propagators as compared to predictions of the RGZ approach.
Our data for the SU(2) ghost propagator $G(p^2)$ in $d= 3,4$ and $2$ have
been compared first to the ``direct'' one-loop formulae, using
the parameters from the gluon-propagator fits reported in
\cite{Cucchieri:2011ig}
and a fixed (theoretical) value for the bare coupling $g^2$.
This comparison is shown
in Figs.\ \ref{fig:gren}, \ref{fig:gren4d} and \ref{fig:gren2d}
respectively for $d = 3, 4$ and 2. In all cases we show the
data (multiplicatively) normalized to
$1/\omu^2$ for $p=\omu=2.5 \,\text{GeV}$. The proposed (one-loop) behavior
is shown with the same normalization on the left side of the figures and, in
all cases, there is a good description in the UV limit.
On the right side of these figures, we have fixed the analytic form to match
the numerical result at
the smallest nonzero (lattice) momentum for the considered lattice volume
and $\beta$ value, i.e.\ we plot a global rescaling of the one-loop prediction.
We find that a good description of the IR region is obtained in
$3d$ and $4d$, confirming that the IR behavior of $G(p^2)$ in these cases is
simply enhanced by a factor with respect to the UV one.
On the contrary, such a rescaling does not hold in $d=2$, since
$G(p^2)$ is IR-enhanced in this case. This difference in IR behavior is
clearly seen in Fig.\ \ref{fig:sigma-comp}, where we show the
Gribov ghost form factor $\sigma(p^2)$ [see Eq.\
(\ref{eq:sigmabis})] for the lattice volumes $V = 320^2, 320^3$
and $128^4$ (respectively the largest volumes for each dimension $d$)
as a function of the unimproved momenta [see Eq.\ (\ref{eq:punimpr})].
In particular, one clearly sees that in
the $3d$ and $4d$ cases the Gribov ghost form factor becomes almost
constant at small momenta.

Next, we have shown the data as compared to the fitted one-loop prediction,
i.e.\ we have used the same parameters as above, but fitting the value of
the bare coupling $g^2$ to the data. A good description is obtained, with
reasonable values of $\chi^2/dof$ (respectively 4.5, 1.7 and 1.6 for
$d = 3,4$ and 2), as seen in the left-hand side of
Figs.\ \ref{fig:gfit}, \ref{fig:gfit4d} and \ref{fig:gfit2d}
respectively for $d = 3, 4$ and 2. We note that an
even better description (respectively with $\chi^2/dof$ of 2.9, 0.75 and
1.2 for $d = 3,4$ and 2) is obtained by fitting the function in
Eq.\ (\ref{eq:FG}) for $d=3,4$ and in Eq.\ (\ref{eq:FG2d}) for $d=2$,
as can be seen in the plots on the right in the same figures.
The fact that one can describe well the whole range of data by using
the analytic prediction for $G(p^2)$ with a fitted value for $g^2$ is
an indication of the importance of having a one-parameter family of
solutions for the propagators in SU($N_c$) Yang-Mills theories
(see Section \ref{sec:intro}). We remark that the ratio between the fitted
value of $g^2$ and the fixed theoretical value is found to be quite
large\footnote{A large value of $g^2$ has also been obtained in Refs.\
\cite{Tissier:2010ts,Tissier:2011ey} by fitting $4d$ and $3d$ lattice
data for the gluon and the ghost propagators to analytic predictions
obtained using a particular case of the Curci-Ferrari model
\cite{Curci:1976bt}.}
in all three cases considered, namely it is about 7.7, 1.9 and 26
respectively for $d=3, 4$ and 2.

Finally, we have isolated the difference between the nonperturbative data
and the one-loop results, by defining the function $W(p^2)$ in
Eq.\ (\ref{eq:G3dW}). As seen in Figs.\ \ref{fig:sigmadif},
\ref{fig:sigmadif4d} and \ref{fig:sigmadif2d}
respectively for $d = 3, 4$ and 2, this difference is
small in the UV region and grows in the IR region. Moreover,
the behavior of $W(p^2)$ is very similar in the three cases and, indeed,
it may be reasonably well parametrized by a simple function
of the momentum [see Eq.\ (\ref{eq:W})]. This supports a unified
explanation for the inaccuracy of the one-loop predictions in the IR
region for the three cases. By considering the similar studies carried out
in Refs.\ \cite{Dudal:2012zx,Aguilar:2013xqa,Aguilar:2014rva} in $d=4$,
it is reasonable to assume that the use of a fully nonperturbative gluon
propagator $D(p^2)$ in the one-loop analysis for $G(p^2)$ is not
sufficient if one does not also use an improved ghost-gluon vertex.
A detailed study of this vertex will be presented elsewhere \cite{inprep}.


\vskip 1cm

\section*{Acknowledgments}

A.C.\ and T.M.\ acknowledge partial support from CNPq, while N.V.\
acknowledges the financial support from the Research Foundation-Flanders
(FWO).
Fits have been done using {\tt gnuplot 4.6} and
{\tt python 2.7.6}. In particular, in the $2d$ case, when
we need to consider the incomplete Beta function with
complex argument we used a {\tt python} code employing
{\tt optimize.leastsq} and {\tt mpmath.betainc}. We have
checked that, when the fits could be done both with gnuplot
and our {\tt python} code, we obtained the same results.


\end{document}